\documentclass[useAMS,usenatbib]{mn2e}
\usepackage{graphicx}
\usepackage{amsmath}
\usepackage{array}
\usepackage{multicol}
\usepackage{multirow}
\usepackage{lscape}
\usepackage{rotating}
\usepackage{url}
\usepackage{color}
\usepackage[colorlinks=true,
            linkcolor=blue,
            urlcolor=blue,
            citecolor=black]{hyperref}
\usepackage{longtable}
\usepackage{rotating}
\usepackage{lscape}

\title[Hubble Diagrams of Relativistic SNe Ic-BL]{Hubble diagrams of relativistic broad-lined type Ic supernovae}

\author[Cano et al.]{\noindent Zach Cano$^{1}$\thanks{zewcano@gmail.com}, P\'all Jakobsson$^{1}$ \& Oli P\'all Geirsson$^{2}$  \\
\noindent $^{1}$Centre for Astrophysics and Cosmology, Science Institute,  Dunhagi 5, University of Iceland, 107 Reykjavik, Iceland.\\
\noindent $^{2}$Department of Mathematics, School of Engineering and Natural Sciences, Dunhagi 5, University of Iceland, 107 Reykjavik, Iceland.\\
}

\date{Accepted XXX. Received YYY; in original form ZZZ}

\pubyear{2015}

\begin{document}
\label{firstpage}
\pagerange{\pageref{firstpage}--\pageref{lastpage}}
\maketitle

\begin{abstract}
This paper is a demonstration of how relativistic, broad-lined type Ic supernovae (SNe IcBL) can be used to constrain a value of the Hubble constant in the local universe ($H_{0}$).  Included in our sample of SNe IcBL are the SNe associated with long-duration $\gamma$-ray bursts (GRB-SNe), as well as local relativistic SNe IcBL that are not associated with GRBs.   Building upon recent results that found a luminosity--stretch relationship for a sample of nine GRB-SNe, we demonstrate that GRB-SNe and SNe IcBL have statistically significant luminosity--decline relationships in $BVR$ filters at the $p=0.02$ confidence level.  Conversely, we show that SNe Ib, Ic and IIb do not have statistically significant luminosity--decline relationships.  Two of the relativistic SNe IcBL in our sample have independent distance measurements (SN~2009bb and SN~2012ap) that we used to constrain a weighted-average value of ${H}_{0,\rm w}=82.5\pm8.1$~km~s$^{-1}$~Mpc$^{-1}$.  This value is 1$\sigma$ greater than that obtained using SNe Ia, and 2$\sigma$ larger than that determined by Planck.  This difference can be attributed to large peculiar motions of the host galaxies of the two SNe IcBL, which are members of galaxy groups.  When determining a value of $H_{0}$ using SNe Ib, Ic and IIb, we found an average value of $H_{0}$ that has a standard deviation of order 20--40~km~s$^{-1}$~Mpc$^{-1}$, which demonstrates that these SNe are poor cosmological candles.  With the launch of the next generation of space telescopes, SNe IcBL, and in particular GRB-SNe, have the potential to constrain constrain the Hubble diagram up to a redshift of $z=3-5$.
\end{abstract}

\begin{keywords}
keyword1 -- keyword2 -- keyword3
\end{keywords}



\section{Introduction}
It was recently shown by Cano (2014; C14 here after) and Li \& Hjorth (2014) that $\gamma$-ray burst supernovae (GRB-SNe), i.e. those supernovae that accompany long-duration GRBs (LGRBs), have the potential to be used as standardizable candles.  In C14, it was shown that a statistically significant correlation was present in the luminosity ($k$) and stretch ($s$) factors (i.e. a luminosity--stretch relationship) of nine GRB-SNe relative to a template supernova: SN~1998bw, which was associated with GRB~980425 (e.g. Galama et al. 1998; Patat et al. 2001).  In the same paper, C14 also showed that relativistic broad-lined type IcBL SN~2009bb, which was not associated with a GRB, also followed the same luminosity--stretch relationship as the GRB-SNe. In a complementary, but independent analysis, a statistically significant luminosity--decline relationship was seen in the K-corrected\footnote{i.e. the correction that arises from observing events occurring at vast cosmological distances, where the observer-frame light arises from bluer rest-frame light that has been redshifted as it traverses through an expanding universe.} $V$-band light curves (LCs) of eight GRB-SNe investigated by Li \& Hjorth (2014).  These results demonstrated that relativistic SNe IcBL, especially GRB-SNe, have the potential to be used as standardizable candles.

In the cosmic distance ladder, type Ia SNe occupy one of the top-most rungs, helping to determine the distance to objects up to redshifts of almost $z=2$ (Jones et al. 2013).  Before the luminosity--decline relationship was observed for nine nearby ($\log(cz)<4.1$) SNe Ia by Phillips (1993), several papers had estimated the Hubble constant ($H_{0}$) by assuming a single peak luminosity for SNe Ia in a given filter (e.g. Kowal 1968; Branch \& Tammann 1992; Sandage \& Tammann 1993).  The luminosity--decline relationship was used thereafter to calibrate SN data in Hubble diagrams to estimate $H_{0}$  (e.g. Hamuy et al., 1995;  Perlmutter et al. 1997, who also estimated the mass budget of the universe).  The use of SNe Ia as standardizable candles reached a crescendo in the late 1990s and early 2000s, when Riess et al. (1998), Perlmutter et al. (1999) and Freedmann et al. (2001) constrained not only the Hubble constant, but also the mass and energy content of the universe, and demonstrated for the first time the existence of ``dark energy'', which is responsible for the accelerated expansion of the cosmos, and which makes up almost three-quarters of everything in the universe.

In this paper, we analysed a sample of relativistic SNe IcBL, including the GRB-SN sample in C14, to first search for luminosity--decline relationships and to then test if they can also be used as cosmological probes in much the same fashion as SNe Ia, and hence provide an estimate of the Hubble constant.  Moreover, we were interested if GRB-SNe, and in turn all relativistic SNe IcBL can obtain as precise a value of $H_{0}$ as other methods.  Crucially, linear distances have been determined for two SNe IcBL in our sample (SN~2009bb \& SN~2012ap), which have allowed us to use them to determine a true value of $H_{0}$.  Moreover, we demonstrate that GRB-SNe and SNe IcBL can be combined into a single sample of ``relativistic SNe'', and together they have a similar root-mean squared (rms, denoted in this paper as $\sigma$) scatter as SNe Ia in their respective Hubble diagrams.  This result hints at their potential to be used in a similar fashion as SNe Ia to constrain not only the value of $H_{0}$ in the local Hubble flow, but cosmological models over almost all of cosmic time.

In Section \ref{sec:Relativistic_SNe} we discuss the physical motivation for combining the local, relativistic SNe IcBL with the GRB-SNe sample to create a single ``relativistic SN'' sample that is used throughout our analysis.  In Section \ref{sec:method} we describe how we assembled our SN sample, including the selection criteria, and how we modelled the K-corrected LCs.  In this section we also describe our compilation of published independent linear distances to various SNe Ib, Ic and IIb in our sample (Table \ref{table:distances}).  In Section \ref{sec:analysis} we describe the two fitting techniques used in this paper to determine the luminosity--decline relationships of our SNe (Section \ref{sec:lumdecline}), and the offsets of the Hubble ridge line to our datasets (Section \ref{sec:hubble_diagrams}) which leads to our determination of $H_{0}$ (Section \ref{sec:Hubble_constant}).  In Section \ref{sec:discussion} we discuss various limitations of our results, including their redshift ranges,  limitations of the fitting methods, and sample sizes.  We present our conclusions in Section \ref{sec:Conclusions}.  In Appendix \ref{sec:Appendix_methods}, we present details of our fitting methods (a Monte-Carlo linear-least squares bootstrap approach, and a Bayesian inference method).  Then in Appendix \ref{sec:Appendix_Ibc}, we present luminosity--decline relationships (or more precisely, the lack thereof) of the SNe Ib, Ic, IIb and SLSN-Ic in our sample, as well as Hubble diagrams constructed for these SNe.  We also discuss the curious case of type Ic SN~1994I, which appears to follow the same luminosity--decline relationship of the relativistic SNe IcBL, but is inconsistent in their combined Hubble diagram.

\section{Relativistic Supernovae}
\label{sec:Relativistic_SNe}

Radio observations of non-thermal material moving at mildly relativistic velocities were obtained of type IcBL SN~2009bb (Soderberg et al. 2010) and SN~2012ap (Chakraborti et al. 2015; Margutti et al. 2014).  The only way to explain these observations was if the outflow had been generated by a powerful central engine -- thus their ``engine-driven'' designation (e.g. Drout et al. 2011).  This explosion scenario is in contrast to the more general neutrino-driven, hydrodynamical core-collapse mechanism used to explain the majority of all core-collapse SNe (see the reviews by, among others, Woosley \& Weaver 1986; Woosley \& Janka 2005; Janka 2012).  The central engine driving the explosion of GRB-SNe is thought to either be an accreting black hole (Woosley 1993; MacFadyen \& Woosley 1999; MacFadyen et al. 2001; Zhang et al. 2004), or a highly magnetised ($\textbf{B} \sim 10^{14-15}$~G), rapidly rotating neutron star, i.e. a magnetar (Usov 1992; Duncan \& Thompson 1992; Wheeler et al. 2000; Thompson et al. 2004; Metzger et al. 2015).  In both scenarios a bipolar jet is launched, either driven by neutrino annihilation preferentially directed towards the polar axes (MacFadyen \& Woosley 1999), or activated by the Blandford--Znajek mechanism (e.g. Komissarov \& Barkov 2009), at ultra-relativistic velocities.

Observational evidence for the presence of both types of central engines have accumulated over the years.  The distribution of $\gamma$-ray pulses measured by CGRO-BATSE (Meegan et al. 1992), $\emph{Swift}$-BAT (Gehrels et al. 2004), and \emph{Fermi}-LAT (Atwood et al. 2009) was shown to be consistent with predictions made by GRBs arising from accereting black holes (Bromberg et al. 2012).  Analytical models (e.g. Zhang \& M\'esz\'aros 2001)  that describe emission arising from a magnetar central engine have been used to explain optical and X-ray observations of both short-duration GRBs (e.g. Rowlinson et al. 2013), LGRBs (Cano et al. 2014; Bernardini et al. 2014; De Pasquale et al. 2015), and ultra-long duration GRBs (Greiner et al. 2015; Metzger et al. 2015; Cano, Johansson \& Maeda 2016).  Moreover, it is also expected that magnetar central engines may be the dominant source of emission powering the light-curves (LCs) of superluminous supernovae (SLSNe); e.g. Kasen \& Bildsten (2010); Woosley (2010); Barkov \& Komissarov (2011); Chatzopoulos et al. (2011); Inserra et al. (2013); Nicholl et al. (2013).

Convincing evidence for both types of central engines is mounting, which appear to power a plethora of highly energetic events, such as very bright SNe and GRBs.  Additionally, there are strong physical motivations for supposing that GRB-SNe and relativistic SNe IcBL\footnote{Not all SNe IcBL are thought to be relativistic SNe, with one clear example being SN~2002ap: in this event, large ejecta velocities were measured ($v \approx 15,000$ km~s$^{-1}$; Gal-Yam et al. 2002), but it is thought that the explosion mechanism was neutrino-driven rather than engine-driven; see below.} arise from similar physical means.  Based on a statistical analysis of photometric LCs of the largest sample of stripped-envelope, core-collapse SNe (i.e. SNe Ib, Ic, IcBL and GRB-SNe) ever considered, Cano (2013; C13 hereafter) demonstrated that GRB-SNe are the most energetic SN Ibc subtype, and are, on average, 10 times more energetic than SNe Ib and Ic (and hence their ``hypernova'' designation; Iwamoto et al. 1998).  SNe IcBL are, on average, five times more energetic than SNe Ib and Ic, and  half as energetic as GRB-SNe.  The average ejecta mass and nickel content therein of SNe IcBL and GRB-SNe was commensurate, and roughly two-times greater than the average SN Ib and Ic.  Based on this, C13 surmised that SNe IcBL and GRB-SNe both arise from engine-driven explosions, where the former are powered by less-energetic central engine than for GRB-SNe.  The weaker engines powering SNe IcBL are due to either more mass being lost in the former events (and hence angular momentum to power the central engine), or the former possessing larger progenitor radii that chokes and then extinguishes the launched jet before it breaks out into space, as is expected for GRB-SNe.  A similar conclusion was also drawn by Modjaz et al. (2015) based on a statistical analysis of almost 400 spectra of 37 SNe Ic, IcBL and GRB-SNe.

Adding complexity to this physical scenario is the presence of low-luminosity $\gamma$-ray bursts ($ll$GRBs), and their accompanying SNe: e.g. GRB~980425 \& SN~1998bw (Galama et al. 1998; Patat et al. 2001); GRB~060218 \& SN~2006aj (Pian et al. 2006; Mazzali et al. 2006; Soderberg et al. 2006) and GRB~100316D \& SN~2010bh (Cano et al. 2011a; Olivares et al. 2012; Bufano et al. 2013).  In terms of high-energy $\gamma$-ray emission, these events are under-luminous by a factor of a hundred to ten thousand relative a typical LGRB: with regards to their isotropic $\gamma$-ray emission, the former have $E_{\rm iso, \gamma} = 10^{46} - 10^{50}$ erg, while the latter have $E_{\rm iso, \gamma} = 10^{51} - 10^{53}$ erg.  The different magnitudes of $\gamma$-ray energetics, as well as the difference in the pulse duration, shape, and spectral properties of $ll$GRBs relative to LGRBs, hints at different physical processes generating the $\gamma$-ray emission in these events.  It has been suggested by many authors (e.g. Campana et al. 2006; Waxman, M\'esz\'aros \& Campana 2007; Cano et al. 2011; Bromberg, Nakar \& Piran 2011; Nakar \& Sari 2012; Margutti et al. 2015; Nakar 2015), but refuted by others (e.g. Ghisellini et al. 2007; Friis \& Watson 2013) that the high-energy emission in $ll$GRBs arises from a relativistic shock-breakout (SBO).   The predicted $\gamma$-ray LC of a SBO event is smooth and non-variable, and has a (soft) $\gamma$-ray energy release of order $10^{44} - 10^{46}$ erg, over a duration of seconds to minutes (Nakar \& Sari 2012).  Such predictions are consistent with the $\gamma$-ray properties of $ll$GRBs, and are clearly inconsistent with those of LGRBs.  

However, the SNe that accompany LGRBs are identical to those of $ll$GRBs: C13 showed that the bolometric properties (kinetic energy, ejecta mass and nickel content in the ejecta) of the SNe associated with $ll$GRBs and LGRBs are statistically indistinguishable.  Furthermore, when the relativistic (jet/collimated outflow) and non-relativistic (the SN) components of LGRBs, $ll$GRBs and relativistic SNe IcBL are considered, further clues are revealed.  Soderberg et al. (2010); Margutti et al. (2013); Margutti et al. (2014); and Cano et al. (2015) have shown that when you consider the kinetic energy ($E_{\rm K}$) and velocity (in the form of $\Gamma \beta$; where the former is the Lorentz factor of the relativistic outflow, and $\beta = v/c$) at the same post-explosion epoch (chosen as +1 day in the rest-frame), that the power-law index ($\alpha$) linking the relativistic and non-relativistic component, $E_{\rm K} \propto (\Gamma\beta)^{-\alpha}$, depends on the nature of the explosion mechanism.  Using Fig. 13 in Cano et al. (2015) as an example, the expected power-law index of a neutrino-driven, hydrodynamical collapse is very steep, and has a value of $\alpha \approx 5$ (Sakurai 1960; Matzner \& McKee 1999; Tan et al. 2001); whereas strongly collimated, engine-driven events such as GRBs have much shallower values of $\alpha = 0.4$ (Lazzati et al. 2012).  Interestingly, both $ll$GRBs and relativistic type IcBL SNe 2009bb and 2012ap occupy the same position in the $E_{\rm K}$--$\Gamma \beta$ plane, and have an index with an intermediate value of $\alpha \approx 2.4$.  

Thus several lines of observational and physical arguments, presented both here and in the literature, suggest that $ll$GRB-SNe and relativistic SNe IcBL\footnote{In this context, SN~2002ap has a power-law index of $\alpha \approx 5.0$, strongly suggesting it is not an engine-driven SN.} arise from similar physical processes, and $ll$GRB-SNe, LGRB-SNe and non-GRB SNe IcBL have similar physical and observational properties.  If this is indeed the case, which the evidence appears to support, then it may be appropriate to combine GRB-SNe and non-GRB relativistic SNe IcBL into a single class of relativistic SNe IcBL, as has been done in the literature for statistical analyses of SNe Ibc (e.g. Richardson et al. 2002; Richardson et al. 2006; Cano 2013; Lyman et al. 2014; Bianco et al. 2014; Modjaz et al. 2014; Richardson et al. 2014), and as we have done here.

\section{Method}
\label{sec:method}

\subsection{The Supernova sample}

In this paper we are interested in SNe with redshifts $z\le0.2$.  This includes six GRB-SNe and four SNe IcBL.  In Appendix \ref{sec:Appendix_Ibc}, we also considered 11 SNe Ib, five SNe Ic, eight SNe IIb, and four SLSNe-Ic, making a total of 38 SNe investigated here.  Of these, eight SNe Ib, five SNe Ic, seven SNe IIb and two SNe IcBL have independent distance measurements to its host galaxy that will ultimately be used to estimate the value of $H_{0}$.   

In order for a given SN to be included in our sample, it must have the following information/observations:

\begin{itemize}
 \item Observations in two or more filters that bracket a rest-frame $BVR$ filter(s).
 \item The magnitudes must be host-subtracted (either mathematically or via the image-subtraction technique).
 \item Knowledge of the entire line-of-sight extinction (both local to the SN \& from the Milky Way).
 \item The LC in a given filter must be sampled well enough that an estimate of its peak time, peak magnitude and the $\Delta m_{15}$ parameter\footnote{i.e. the amount the light curve fades in a given filter from peak light to fifteen days later} can be determined.
\end{itemize}

Once SNe with suitable observations were identified in the literature, we undertook the following general steps to obtain decomposed K-corrected LCs of each SN:

\begin{enumerate}
 \item Remove the host contribution.
 \item Correct for foreground extinction.
 \item Convert magnitudes into monochromatic fluxes using zeropoints in Fukugita et al. (1995).
 \item Create observer-frame spectral energy distributions (SEDs) and interpolate to $BVR$~$(1+z)$ wavelengths and extract the flux.
 \item Correct for rest-frame extinction.
\end{enumerate}

In addition to the above steps, the GRB-SN LCs need to be further decomposed in order to isolate the flux coming from the SN itself.  Such a decomposition is not required for $ll$GRB-SNe due to the insignificant, or completely absent, optical afterglow (AG) component.  For every cosmological GRB-SN event, light arises from three sources (Zeh et al. 2004; Ferrero et al. 2006; Cano et al. 2011a; Hjorth 2013): the AG, the accompanying SN, and a constant source of flux from the underlying host galaxy.  In addition to the host-flux removal, the AG contribution was removed by fitting the host-subtracted LCs with an AG model.  The light that powers a GRB AG is expected to be synchrotron in origin, and thus has a power-law dependence in both time and frequency\footnote{$f_{\nu} \propto (t - t_{0})^{-\alpha}\nu^{-\beta}$, where $t_{0}$ is the time at which the GRB triggered by a GRB satellite, and the temporal decay and energy spectral indices are $\alpha$ and $\beta$, respectively.}.  We followed the procedure described in C13, C14, Cano et al. (2014; 2015) to model, and then subtract away the AG contribution, thus leaving flux from only the SN.

\begin{figure}
 \raggedright
 \includegraphics[bb=0 0 576 432, scale=0.46]{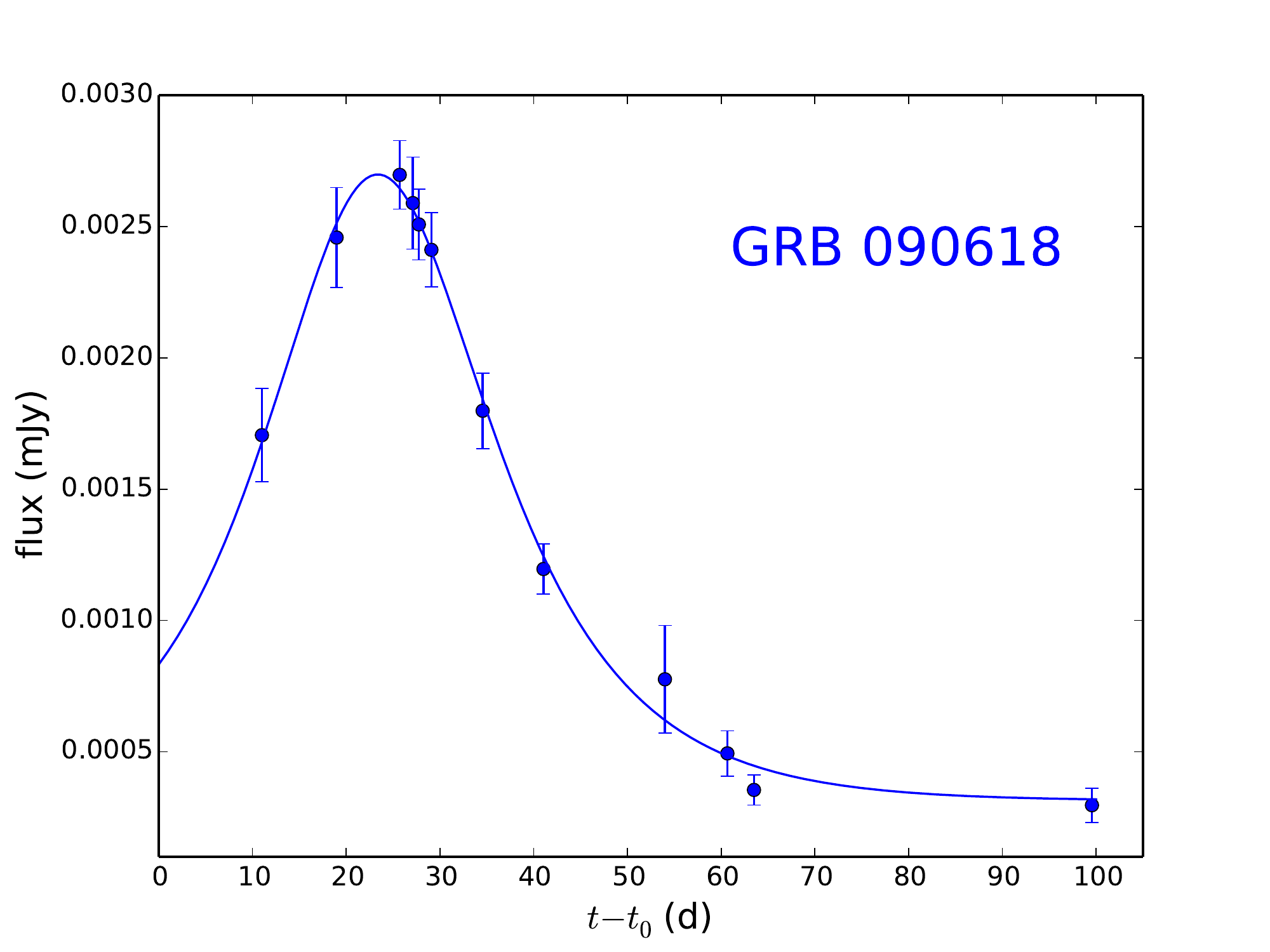}
 \caption{An example of fitting the Bazin function (equation \ref{equ:bazin}, shown as solid blue line) to the interpolated $B$-band (blue points) LC of GRB~090618 from C14 and Cano et al. (2011b).  Note that the parameter $t_{0}$ shown on the $x$-axis is the trigger time of the GRB, rather than one of the free parameters in the Bazin function (equation \ref{equ:bazin}). }
\label{fig:bazin}
\end{figure}

\subsection{Estimation of observable parameters}

Using the observations obtained via the method described in the preceding section, we modelled the resultant SN LCs to determine the (1) peak apparent magnitudes, and (2) $\Delta m_{15}$ for each SN in each filter.  The observables then allowed us to determine for each SN subtype: the (i) the luminosity--decline relationship of a given SN subtype in a given filter, and (ii) the value of $H_{0}$.

We used three different functions to determine these observables: (1) the Bazin function  (Bazin et al. 2011; equation \ref{equ:bazin} in this work; Fig. \ref{fig:bazin}), (2) high-order polynomials, and (3) linear splines.  All functions were fit to the data using \textsc{python} scripts.  A linear-least-squares (LLS) Levenberg-Marquardt algorithm was used via \textsc{scipy.optimize.curve$\_$fit} to fit the Bazin function (eq. \ref{equ:bazin}) to the data, while \textsc{numpy.polyfit} was used to fit the polynomial to the data.  Finally, \textsc{scipy.interpolate.interp1d} was used to fit the linear spline to the data.

\begin{equation}
 f(t) = \Lambda \left[ \frac{\mathrm{exp}\left(\frac{-(t-t_{0})}{\tau_{\rm fall}}\right)}{1+\mathrm{exp}\left(\frac{-(t-t_{0})}{\tau_{\rm rise}}\right)} \right] + \Phi
 \label{equ:bazin}
\end{equation}

The Bazin function possesses fewer free parameters than the empirical model used in Cano et al. (2011a) and C13 to describe the LC shapes of the SNe in our sample.  It has five free parameters: $\Lambda$ is a normalization constant, $\Phi$ is an offset constant, and $\tau_{\rm rise}$ \& $\tau_{\rm fall}$ are the rise and fall times of the SN LC, respectively.  $t_{0}$ is related to the rise and fall times, as well as the time of maximum light by $t_{0} = t_{\rm max} - \tau_{\rm rise} \times \mathrm{ln}\left(\frac{\tau_{\rm fall}}{\tau_{\rm rise} - 1}\right)$.

Note that we adopted the conservative 20\% error estimated by C14 for the peak magnitudes and $\Delta m_{15}$ values of the GRB-SNe.  This conservative error arises from the fact that the isolation of the SN light involves a complicated decomposition technique (as described above), with several sources of error that arise from GRB AG modelling and subtraction, the host-galaxy subtraction, the uncertainties in the extinction (both from sight-lines through the Milky Way, and from the SN's host galaxy), and the SED interpolation.  The 20\% error adopted here is much larger than the systematic error associated with the different fitting functions described above.  We have estimated and propagated errors in our analysis for the remaining SNe in our sample in a similar fashion, which are smaller than those of the GRB-SNe due to the fact that there are no uncertainties associated with the AG modelling and removal.

The final dataset, which includes peak apparent magnitudes and their associated errors, spectroscopic redshifts, line-of-sight extinction, as well as citations to the various published datasets for each SN are displayed in Table \ref{table:photometry}.

\begin{figure*}
 \centering
 \includegraphics[bb=0 0 880 207,scale=0.55,keepaspectratio=true]{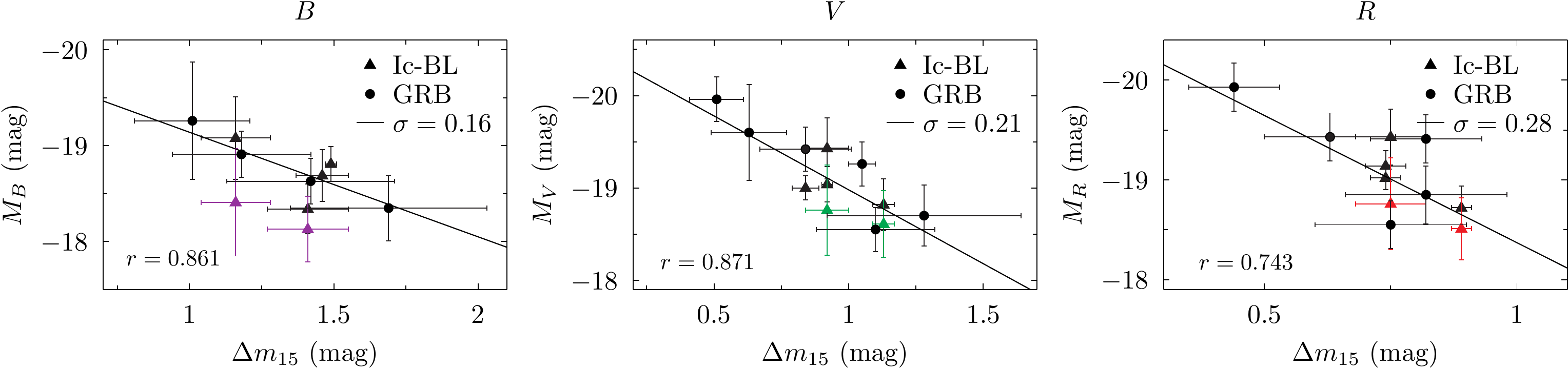}
 \caption{Luminosity--decline relationships of SNe IcBL and GRB-SNe in filters $B$ (purple), $V$ (green) and $R$ (red).  Solid black lines and points correspond to absolute magnitudes calculated for luminosity distances (see Section \ref{sec:linear_distances}), while coloured points and lines correspond to absolute magnitudes calculated for those events where an independent distance measurement(s) have been made to the SN's host galaxy.  The correlation coefficient ($r$) for each dataset is shown (in black and in their respective colours) as well as the best-fitting luminosity--decline relationship (eq. \ref{equ:peak_m15}) determined using the MC--LLS bootstrap method, and the corresponding rms ($\sigma$) of the fitted model.  It is seen that statistically significant correlations are present for the GRB-SNe and combined GRB-SN \& SN IcBL samples (see Table \ref{table:lum_decline} for their corresponding $p$ values).  In comparison (Fig. \ref{fig:lum_decline_all}), SNe Ib, Ic, IIb and SLSNe-Ic (except perhaps in the $R$-band) do not have a luminosity--decline relationship. }
 \label{fig:lum_decline_IcBL}
\end{figure*}

\subsection{Linear distances}
\label{sec:linear_distances}

For many of the SNe in our sample, an estimate of the linear distance to them and/or their host galaxy have been made by one or more authors.  The techniques used to determine these linear distances are: the planetary nebula luminosity function (PNLF), surface-brightness fluctuations (SBF), the diameters of inner ring structures in S(r) galaxies, the expanding photosphere method (EPM), the spectra-fitting expanding atmosphere method (SEAM), the SN II standard candle method (SCM), the Tully-Fisher (TF) method, the brightnesses of the brightest red and blue supergiants, distances determined from the concept of sosie galaxies, using the distances to SNe Ia that have occurred in the same galaxy (e.g. SN 2011bm), and the use of theoretical criteria for the gravitational stability of gaseous disks.  

All of the distance measurements were obtained from the data tabulated in the NASA/IPAC Extragalactic Database (NED\footnote{\url{http://ned.ipac.caltech.edu}}).  Distance measurements for a given SN can vary in precision, with some measurements having small errorbars, while others being poorly constrained and having much larger errorbars.  To account for this fact we computed the weighted average distance ($D_{\rm w}$) to the host galaxy of each SN, which are displayed in Table \ref{table:distances}, along with calculated luminosity distances for each galaxy's redshift\footnote{Determined for a $\Lambda$CDM cosmology constrained by Planck (Planck Collaboration et al. 2013) of $H_{0} = 67.3$ km s$^{-1}$ Mpc$^{-1}$, $\Omega_{\rm M} = 0.315$, $\Omega_{\Lambda} = 0.685$.}.

\section{Analysis \& results}
\label{sec:analysis}

\subsection{Fitting methods}
\label{sec:fitting_methods_short}

In order to derive robust results for the luminosity--decline relationships (Section \ref{sec:lumdecline}), the $y$-intercepts of the Hubble ridge line (Section \ref{sec:hubble_diagrams}) and ultimately the Hubble constant (Section \ref{sec:Hubble_constant}), we employed two different fitting techniques:

\begin{itemize}
 \item A Monte-Carlo, linear-least squares (MC--LLS) bootstrap method.
 \item A Bayesian inference method with nested sampling (MultiNest, Feroz et al. 2009; made usable in \textsc{python} by PyMultiNest, Buchner et al. 2014).\end{itemize}

A detailed description of each fitting method is presented in Appendix \ref{sec:Appendix_methods}.  Both methods (necessarily) account for the errors in both the $x$ and $y$ directions.  Our motivation for using two different fitting methods is to demonstrate the robustness of our results, and that our conclusions are not dependent on the fitting method used.  As GRB-SNe have the potential to be used to constrain cosmological models across almost all of cosmic time (up to $z=$~3--5; see Section \ref{sec:discussion}), demonstrating their applicability to this purpose can only be achieved after they pass stringent testing.

\subsection{Luminosity--decline relationships}
\label{sec:lumdecline}

Figure \ref{fig:lum_decline_IcBL} shows the peak absolute magnitudes of the relativistic SNe IcBL as a function of $\Delta m_{15}$ in filters $BVR$.  The corresponding figure for SNe Ib, Ic, IIb and SLSN-Ic is presented in Appendix \ref{sec:Appendix_Ibc} (Fig. \ref{fig:lum_decline_all}).  We considered absolute magnitudes determined from both linear and luminosity distances (Table \ref{table:distances}).  To each dataset, a linear relation was fitted to the data:

\begin{equation}
 M_{\nu} = m \times \Delta m_{15,\nu} + b
 \label{equ:peak_m15}
\end{equation}

\noindent to determine the slope ($m$) and $y$-intercept ($b$) using the two methods described in Section~\ref{sec:fitting_methods_short} and Appendix \ref{sec:Appendix_methods}.  For each dataset we also calculated Pearson's correlation coefficient ($r$) and the two-point probability of a chance correlation ($p$), where the former are displayed in each subplot in Fig. \ref{fig:lum_decline_IcBL}.  The best-fitting values, as well as the determined values of $r$ and $p$, and sample sizes ($N$), are displayed in Table \ref{table:lum_decline}, where the label $D_{L}$ denotes values determined for luminosity distances, and ``Lin.'' for linear distances.  Note that linear distances exist for SNe Ib, Ic, IIb and IcBL only, and not for any of the GRB-SNe in our sample.

It is seen that a statistically significant luminosity--decline relationship is present for the GRB-SN sample (in all filters), as well as the combined SN IcBL \& GRB-SN sample (in all filters).  The relations in filters $B$ and $V$ of the GRB-SN and combined SN IcBL \& GRB-SN samples are statistically significant at the $p=0.02$ significance level.  However, the relation in $R$ for the GRB-SN sample is not significant, even at the $p=0.1$ level.  The relation in $R$-band of the combined GRB-SN \& SN IcBL sample is statistically significant at the $p=0.05$ level.

The only other statistically significant relation seen for the other SN subtypes (Fig. \ref{fig:lum_decline_all}) was that of the SLSNe-Ic in $R$, which is significant at the $p=0.02$ level.  All other SN subtypes, including the $B$ and $V$ filters of the SLSN-Ic sample, are consistent with having no luminosity--decline relationship.  This conclusion holds for samples derived from both luminosity distances and linear distances.  Our results support the conclusions of Li \& Hjorth (2014) and Inserra \& Smartt (2014), who also found luminosity--decline relationships for GRB-SNe and SLSNe, respectively (see as well the discussion in Section \ref{sec:Appendix_lum_decline_Ibc}).

\subsection{Hubble diagrams}
\label{sec:hubble_diagrams}

\begin{figure*}
 \centering
 \includegraphics[bb=0 0 910 313,scale=0.55,keepaspectratio=true]{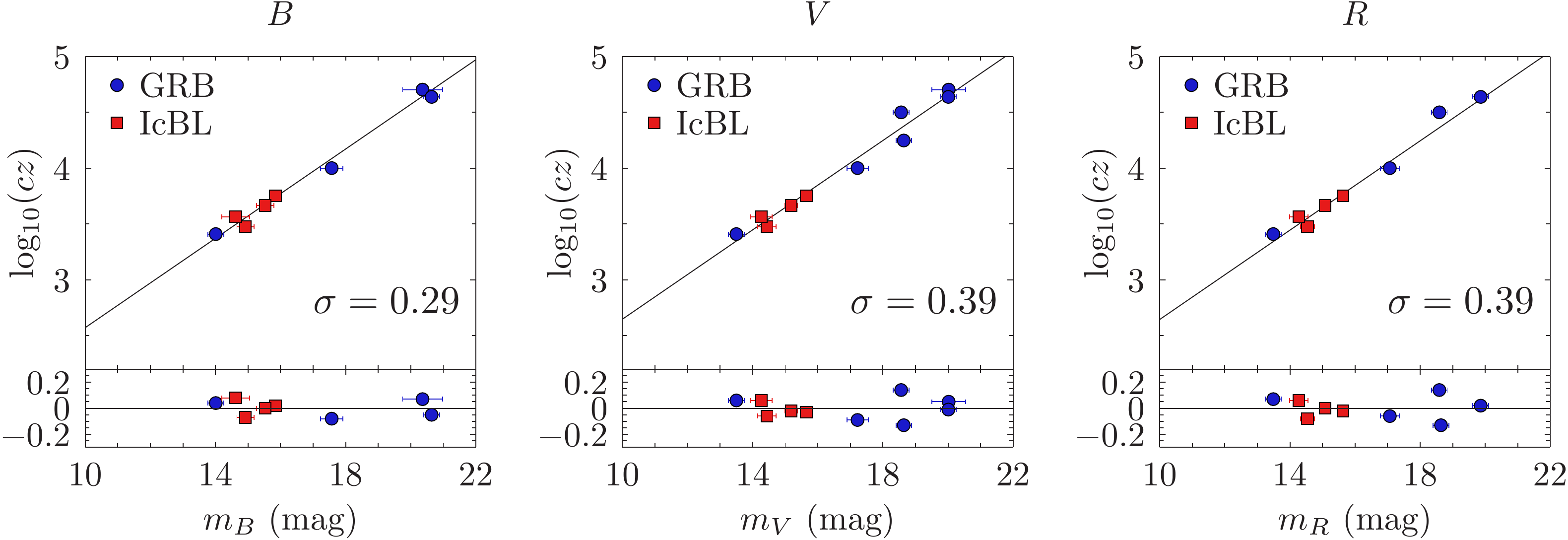}
 \caption{Hubble diagrams of relativistic SNe IcBL in filters $BVR$.  Plotted in each subplot are the uncorrected magnitudes of each subtype and the fitted Hubble ridge line as determined using the MC--LLS bootstrap method.  Also plotted are the rms values ($\sigma$) and residuals of the magnitudes about the ridge line.  In the $B$-band, the amount of scatter in the combined SNe IcBL sample is the same as that for SNe Ia up to $z=0.2$ (not plotted, but values given in Table \ref{table:hubble_offsets_uncorrected}), which is $\sigma \approx 0.3$~mag.}
 \label{fig:hubble_IcBL}
\end{figure*}

Equation \ref{equ:hubbleline} gives the Hubble ridge line (e.g. Tammann et al. 2002; Sandage \& Tammann 1993), which is valid for nearby SNe whose redshift are $z\ll1$.  Given the relative paucity of SNe in our sample, we imposed a limiting redshift for a SN to be included in our sample of $z\le0.2$.  It is important to omit events whose redshifts exceed $z=0.2$ as they are no longer in the local Hubble flow and their inclusion in the fit can result in miscalculating the local value of the Hubble constant.  Our procedure follows the examples used in SNe Ia studies (e.g. Riess et al. 2004; 2007) that impose an upper limit of $z\le0.1$ when fitting to obtain $H_{0}$.  In Section \ref{sec:discussion} we discuss the effect of the two redshift limits ($z\le0.1,0.2$) on the resultant value of $H_{0}$.

\begin{equation}
 \mathrm{log}(cz) = 0.2 \times m_{\rm app} + \delta
 \label{equ:hubbleline}
\end{equation}

As in the previous section, we used both fitting procedures to determine the $y$-intercept ($\delta$) of the Hubble ridge line for magnitudes that were uncorrected (Table \ref{table:hubble_offsets_uncorrected}) for a luminosity--decline relationship.  Figure \ref{fig:hubble_IcBL} shows Hubble diagrams, including ridge lines, of all the relativistic SNe IcBL SNe in our sample whose magnitudes were not corrected for a luminosity--decline relationship.  The corresponding figure that shows the same plot for all of the hydrogen-deficient SNe is shown in Fig. \ref{fig:hubble_diagrams} in Appendix \ref{sec:Appendix_Hubble_Ibc}.  The rms scatter ($\sigma$) and the residuals of the fitted model (both in units of magnitudes) are plotted in each figure.

In Figures \ref{fig:hubble_IcBL} and \ref{fig:hubble_diagrams}, it is seen that the smallest amount of scatter about the Hubble ridge line occurs for the combined GRB-SN and SN IcBL samples ($0.3\le \sigma \le 0.4$ mag).  The SLSN-Ic sample has the second-smallest amount of scatter ($0.3\le \sigma \le 0.5$ mag).  The average rms of the SN Ib sample is $\sigma\approx$0.6 mag, while that of the SN Ic sample ($\sigma=$0.6--0.7 mag) is similar to that of the SNe Ib.  The SNe IIb sample has the largest amount of scatter of the SN subtypes, with $\sigma=$0.8--0.9 mag.  Further discussion regarding the underlying reasons for the different amounts of scatter are presented in Section \ref{sec:discussion}.

\begin{figure}
 \centering
 \includegraphics[bb=0 0 250 172, scale=0.95]{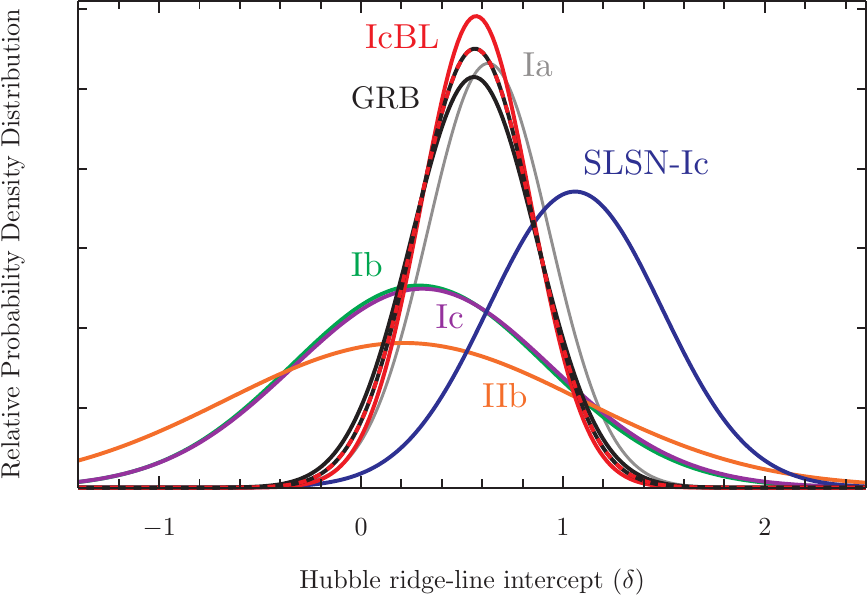}
 \caption{Unit-area probability density functions (pdf) of the sample of hydrogen-deficient SNe in $B$-band, which are not corrected for a luminosity--decline relationship.  The unit-area pdf of each SN subtype is centred on its value of the Hubble ridge-line offset ($\delta$; Table \ref{table:hubble_offsets_uncorrected}) determined using the MC bootstrap method, and which is broadened by the rms ($\sigma$) in that filter (see left-column in Fig. \ref{fig:hubble_diagrams}).  It is seen that the SNe Ia, GRB-SNe and SNe IcBL have the smallest rms values, and thus the narrowest/tallest pdfs.  It is also seen that the value of $\sigma$ of the separate SN IcBL and GRB-SN samples are identical to within their errorbars (Table \ref{table:hubble_offsets_uncorrected}), which hints that they arise from similar types of explosions and may be used together in the same sample (the combined GRB \& IcBL sample is shown as the red \& black dashed pdf).}
 \label{fig:pdfs}
\end{figure}

Another way to visualize the scatter of the magnitudes of each subtype is to plot the relative probability density distribution, comprised of unit-area probability density functions (pdfs), of the entire SN sample, as shown in Figure \ref{fig:pdfs}.  In this figure, a unit-area pdf is plotted that is centred on the offset of each uncorrected SN dataset in the $B$-band, which is broadened by its rms value.  Plotted for comparison is the SN Ia sample from Betoule et al. (2014), which consists of 318 SNe with redshifts $z<0.2$.  The rms of the SN Ia sample is $\sigma=0.30$~mag, which is also shown in Table \ref{table:hubble_offsets_uncorrected}.  It is seen that the SN IcBL and GRB-SN samples, as well as the combined GRB-SN \& SN IcBL samples, have a similar amount of scatter as the SN Ia sample.  It is also interesting to note that the value of the Hubble ridge line intercept is approximately the same between these three SN subtypes, which reflects the fact that they all peak at roughly the same magnitude range ($M_{\rm abs}\sim-18.5$ to $-19.5$ mag).  Similarly, the SLSNe-Ic have a larger Hubble intercept value, which reflects the fact that they peak at even brighter magnitudes, while the SNe Ib, Ic and IIb have smaller intercept values, indicating that their average peak magnitude is fainter than the other SN subtypes considered here.

\subsection{The Hubble constant}
\label{sec:Hubble_constant}

Once the intercept for a given SN subtype in a given filter is determined, equation \ref{equ:hubbleconstant} is used to finally calculate $H_{0}$:

\begin{equation}
 \mathrm{log}(H_{0}) = 0.2 \times M_{\rm abs} + \delta + 5
 \label{equ:hubbleconstant}
\end{equation}

For every SN where an independent distance measurement has been obtain to it, or its host galaxy, it is possible to use it to calculate $H_{0}$.  In our analysis we computed $H_{0}$ for magnitudes that were uncorrected for a luminosity--decline relationship, using the offsets obtained from the two different fitting procedures.  Moreover, we calculated the average ($\bar{H}_{0}$) and weighted average ($H_{0,\rm w}$), as well as the standard deviation of the distribution of values, of the Hubble constant for a given SN subtype in a given filter.  A summary of this analysis is displayed in Table \ref{table:H0_per_SN_per_filter}, while a summary of the collected results is displayed in Table \ref{table:H0_summary}.

The weighted and average values of $H_{0}$ obtained for the combined relativistic SN IcBL dataset, using the MC--LLS bootstrap method are quite similar: $H_{0,\rm w}=82.5\pm8.1$~km~s$^{-1}$~Mpc$^{-1}$, and $\bar{H}_{0}=81.4$~km~s$^{-1}$~Mpc$^{-1}$, with a standard deviation of 6.1 km~s$^{-1}$~Mpc$^{-1}$, which we use as the 1$\sigma$ standard error.  Statistically similar values were found using the Bayesian inference method.  These values are just over 1$\sigma$ larger than that determined using SNe Ia ($H_{0}=73.8\pm2.4$~km~s$^{-1}$~Mpc$^{-1}$, Riess et al. 2011), and almost 2$\sigma$ larger than that determined by Planck ($H_{0}=67.3$~km~s$^{-1}$~Mpc$^{-1}$) and a combined baryon acoustic oscillations (BAO) + CMB + SN Ia analysis ($H_{0}=69.6\pm0.7$~km~s$^{-1}$~Mpc$^{-1}$; Bennett et al. 2014).  Further discussion regarding this apparent discrepancy between $H_{0}$ found using SNe IcBL and other methods is presented in Section \ref{sec:discussion_discrepancy}.

In comparison, the average value of $H_{0}$ obtained using the SNe Ib, SNe Ic and SNe IIb range from $\sim 63 - 123.1$~km~s$^{-1}$~Mpc$^{-1}$, with standard deviations of order 20--40~km~s$^{-1}$~Mpc$^{-1}$.  Such large amounts of scatter in the Hubble diagrams of these SNe (see Fig. \ref{fig:lum_decline_all}), directly results in the large scatter in the derived value of $H_{0}$, and highlights that SNe Ib, Ic and IIb are not very useful cosmological probes.  Further details on the results obtained for these SNe are found in Appendix \ref{sec:Appendix_Hubble_Ibc}.

\section{Discussion}
\label{sec:discussion}

In this paper we demonstrated that relativistic SNe IcBL can be used to measure a weighted-average of $H_{0}$ to a precision of $\sim10$\%.  The uncertainties in the distances to the host galaxies of SN~2009bb ($40.68\pm4.36$~Mpc, 11\% error) and SN~2012ap ($40.37\pm7.38$~Mpc, 18\% error) dominate the error in $H_{0}$.  The small SN IcBL sample size also contributes: the error in the offset of the Hubble ridge line of the SN Ia sample ($N=318$) is $\pm0.001$ in the $B$-band, whereas is is ten times larger ($\pm0.014$) for the SN IcBL sample ($N=8$) in the same filter.  Larger sample sizes will reduce the offset error, but more accurate distance estimations are also required of future SNe IcBL in order to obtain a more constrained value of $H_{0}$.

The scatter seen in the Hubble diagrams for each subtype varied from levels similar to those seen for SNe Ia ($\sigma\approx 0.3$--0.4 mag; e.g. GRB-SNe and SNe IcBL), to much greater scatter ($\sigma\approx 0.8$--0.9 mag for SNe IIb).  The scatter of the SNe Ib, Ic and SLSN-Ic datasets are intermediate between these two extremes.  The resultant uncertainties in $H_{0}$ derived using these SNe vary from about 20--40~km~s$^{-1}$~Mpc$^{-1}$. 

The reason why relativistic SNe IcBL can be used to constrain $H_{0}$ to a relatively precise value arises from their distribution of peak magnitudes.  Historically, SNe II were not reliable suitable standard(izable) candles because they displayed considerable variation in their peak luminosities (with a scatter of $\sim 1$~mag in their Hubble diagrams; e.g. Hamuy \& Pinto 2002).  The degree of scatter observed here for SNe IIb imply that they have a similar distribution of peak magnitudes as all SNe II, while  SNe Ib and Ic have a slightly less amount of scatter, but it is not small enough to allow them to be useful standard candles.

Richardson et al. (2014) investigated both the bias-corrected and volume-limited distributions of the peak magnitudes of all types of SNe, finding that SNe Ia displayed the lowest amount of scatter (standard deviation of $\approx 0.5$~mag) of all the SN subtypes.  SNe IIn had the largest amount of scatter (standard deviation of $\approx 1.5$~mag), with the remaining subtypes (SNe Ib, Ic, IIb, IIL and IIP) all having roughly the same amount of scatter (standard deviations of $\approx 1$~mag).  In a previous paper by the same lead author, Richardson et al. (2006) split their entire SN Ic sample into ``normal'' and ``bright'' subgroups, where the scatter of each decreased from a standard deviation of 0.9 mag down to 0.4 and 0.5 mag, respectively.   When comparing with the subgroups investigated by Richardson et al. (2006), clearly our sample here has the same amount of scatter as theirs, and the scatter of the peak magnitudes of the different SN subtypes in the Hubble diagrams can be readily explained by the distribution of their peak magnitudes. 

Interestingly, type IcBL SN~2003jd and SN~2007ru follow the luminosity--decline relationship of the other relativistic SNe IcBL, while type IcBL SN~2002ap does not, nor do type Ic SN~2004aw and SN~2007gr.  For SN~2003jd, Valenti et al. (2008) investigated whether this SN could have been associated with a GRB or $ll$GRB, perhaps one seen off-axis.  The main argument presented by Valenti et al. (2008) against a GRB association was based on the non-detection of this event at radio wavelengths, thus negating an engine-driven origin.  The earliest radio observation of SN~2003jd at $\sim 10$~d post-peak resulted in an upper limit to the radio flux that implied a radio luminosity of roughly a factor of ten less than SN~2006aj at a comparable epoch.  In Fig. 13 of Cano et al. (2015), this would place it in a region intermediate between ordinary SNe Ibc and the relativistic SNe IcBL (though we must consider the caveat that times of the energetics in Fig. 13 are taken at +1~d from explosion).  As such, one cannot rule out an engine-driven origin for this event.  For SN~2007ru, no radio observations were obtained by Sahu et al. (2009).  Nevertheless, it should be pointed out that the inclusion/exclusion of these two events in the relativistic SN IcBL sample has no qualitative consequence on our results; their inclusion/omission may have some quantitative effects in the derived values of the Hubble ridge line $y$-intercept values, but these effects will be minimal and change the value of $H_{0}$ by only 1--3~km~s$^{-1}$~Mpc$^{-1}$.

\subsection{Discrepancy in $H_{0}$ between SNe Ia and IcBL}
\label{sec:discussion_discrepancy}

While we have demonstrated that relativistic SNe IcBL can be used to determine a well-constrained value of $H_{0}$, the biggest surprise is our derived value of ${H}_{0,\rm w}=82.5\pm8.1$~km~s$^{-1}$~Mpc$^{-1}$.  As noted previously, this is 1$\sigma$ and 2$\sigma$ greater than the values found using just SNe Ia, and combined datasets (SNe Ia, CMB maps and BAOs).  

The cause for this difference it not immediately obvious, however there are a few reasons why we could have measured a larger value here.  The first may be that we have underestimated the amount of line-of-sight extinction towards these events, thus causing them to be artificially fainter, and hence leading us to derive a larger value of $H_{0}$.  To investigate this, we turned to the observational photometric datasets of SN~2009bb and SN~2012ap, and used the empirically derived relation between $V-R$ at +10~d in both $V$ and $R$ determined by Drout et al. (2011).  Correcting the photometry by the extinction values calculated by Levesque et al. (2010) and Liu et al. (2015) for SN~2009bb and SN~2012ap, respectively, we then checked whether the $V-R$ colours were in excess of the relations in Drout et al. (2011).  In both cases, the excess extinction was found to be negligible, and we conclude that the extinction of these two SNe has been properly accounted for.

Another reason may arise from incorrect, or very uncertain, distance measurements being made of the host galaxies of SN~2009bb (NGC 3278) and SN~2012ap (NGC 1729), as listed in NED.  For example, the calculated distance to NGC~3278 is $40.7\pm4.4$~Mpc.  For distances of 40 and 45~Mpc, the difference in the calculated distance moduli is 0.3~mag.  If the distance is at the upper limit of this value, we then derive $H_{0} \approx 77$~km~s$^{-1}$~Mpc$^{-1}$, which is no longer in tension with that determine via SNe Ia.  Both of these galaxies have TF distances determined by Theureau et al. (2007) and Springob et al. (2009), respectively.  The distances presented by Springob et al. (2009) are corrected for Malmquist bias, while those of Theureau et al. (2007) are not.  The distances to these galaxies were determined \emph{en masse}, in samples of $>$2000 and 3126 objects, respectively.  There is no reason to suspect that the distance measurements are somehow erroneous, and we are forced to assume that they can be regarded as being correctly determined.

Another possibility may arise from the fact that these galaxies are not in the local Hubble flow.  If they have large peculiar velocities, perhaps arising from their position in a galaxy cluster, then the value of $H_{0}$ derived from using these SN will not be a suitable representation of the ``true'' value of $H_{0}$.  First, it appears that NGC~3278 may be part of a galaxy group (e.g. Lauberts 1982): its nearest neighbour is NGC~3276, and they have an angular separation of 4.8$''$.  For a standard cosmology, this amounts to a projected distance of 60--65 kpc; i.e. close enough to be interacting gravitationally, which may be affecting their motions relative to the local Hubble flow.  For NGC~1729, its measured linear distance ($40.4\pm7.4$~Mpc) and luminosity distance ($\approx 55$~Mpc) already hint that something peculiar might be at play.  Indeed this galaxy is also a group member (e.g. Garcia 1993), and the comparison of the linear and luminosity distances indirectly suggest that gravitational effects are also affecting the motion of this galaxy relative to the local Hubble flow.

\subsection{Effect of the chosen redshift upper limit}

As discussed in Section \ref{sec:hubble_diagrams}, the Hubble ridge line (eq. \ref{equ:hubbleline}) is only valid for small redshifts ($z\ll1$).  In this work we imposed an upper limit of $z\le0.2$.  Other works that used SNe Ia as cosmological probes (e.g. Riess et al. 2004, 2007) imposed an upper limit of $z\le0.1$.  These works considered many events, and thus imposing this very strict upper limit means that they only probed the very local Hubble flow that was not influenced at all by the energy and mass densities of the universe.  In this work we reached a compromise of $z\le0.2$ based on the modest sample sizes being considered here.  However we must still ask whether this looser redshift limit will affect our final calculated value of $H_{0}$.

Using the SN Ia sample in Betoule et al. (2014), we fit the ridge line to two samples: (1) $z\le0.2$ and (2) $z\le0.1$.  In case (1) we found $\delta = 0.631$, while in case (2) we found $\delta = 0.643$.  The difference between the two samples amounts to $\Delta \delta = 0.012$.  The effect of the slightly different value of the offset on the value of $H_{0}$ determined for a given SN will have different effects on different SNe.  For example, in eq. \ref{equ:hubbleconstant} we can see that $H_{0}$ is directly proportional to $10^{0.2\times M_{\rm abs}}$, therefore fainter SNe will be more affected by a different value of $\delta$ than brighter SNe\footnote{Take as an example two SNe Ib with absolute magnitudes of -18 and -18.5, as well as eq. \ref{equ:hubbleconstant}.  Multiplying each by 0.2, and adding 5, gives 1.4 and 1.3, respectively.  If we assume an offset of $\delta=0.400$, we find $H_{0}=63.1,50.1$~km~s$^{-1}$~Mpc$^{-1}$ , respectively.  If instead $\delta=0.412$, we find $H_{0}=64.9,51.5$~km~s$^{-1}$~Mpc$^{-1}$, respectively.  The difference amounts to $\Delta H_{0}=1.8,1.4$~km~s$^{-1}$~Mpc$^{-1}$, respectively.  Therefore, the calculated value of $H_{0}$ was less effected in the brighter event.}.  In general, the effect of the looser redshift limit will amount to a difference of 1--4~km~s$^{-1}$~Mpc$^{-1}$ in the final value of $H_{0}$, which is smaller than the errors derived here that are based on the standard deviations calculated for each dataset (which we consider to be 1$\sigma$ errors on the calculated value of $H_{0}$).

\subsection{Limitations of the fitting methods}

We used two fitting procedures in our analysis: a MC--LLS bootstrap method and a Bayesian inference method.  We tested both of the fitting methods on an ``ideal'' dataset, the $z\le0.2$ SN Ia sample ($N=318$) from Betoule et al. (2014), to determine the offset of the Hubble ridge line.  As seen in Table \ref{table:hubble_offsets_uncorrected}, the value of the intercept agrees to three decimal places ($\delta = 0.631$).  When the smaller ($N=152$) SN Ia dataset (for redshifts $z\le0.1$) was considered, they also returned the same value for the offset to three decimal places ($\delta = 0.643$).  Thus, when dealing with large datasets, each method returns the same answer.

However, their performance on smaller datasets varied depending on how correlated the samples were.  For example, for datasets that were already quite strongly correlated, such as the uncorrected GRB-SN and combined GRB-SN \& SN IcBL samples, the values of the offsets of the Hubble ridge line were consistent to within the errorbars determined by each method.  The same could not be said for datasets that were not well correlated, such as the uncorrected SN IIb samples.  In these cases, the Bayesian inference method was not as effective in determining the offset, as demonstrated by their larger rms values (Table \ref{table:hubble_offsets_uncorrected}). 

\subsection{Effect of small-sample sizes}

In this paper we demonstrated that the amount of scatter in the uncorrected Hubble diagrams of GRB-SNe, SNe IcBL and the combined GRB-SN \& SN IcBL datasets was approximately the same as that seen for the SNe Ia sample from Betoule et al. (2014).  This is clearly seen in the pdfs in Fig. \ref{fig:pdfs}.  However, we are forced to question if the perceived low values of the scatter of the relativistic SN samples arises due to their small sample sizes -- i.e. if these sample sizes were the same as the SNe Ia, would they still exhibit a low amount of scatter, or would the rms values increase?

To try and address this uncertainty, we performed an analysis whereby we randomly selected (using Monte-Carlo sampling) events from the SN Ia sample over a range of sample sizes ($N=$~4--15) and fit the ridge line to the smaller dataset, and calculated the rms associated with the fit.  In each fit, the rms value of the small, randomly selected SN Ia sample were in the range $\sigma=0.30$--0.35 mag, which is fully consistent with the rms scatter calculated for the full SN Ia sample.

If the range of scatter in the smaller SN Ia samples were smaller, then we could conclude that the small scatter seen for the combined GRB-SN \& SN IcBL sample was merely a consequence of the small sample sizes used here, and that the actual scatter of the entire relativistic SN population is intrinsically higher.  Instead, the fact that we found similar values for the rms scatter of the smaller SN Ia samples implies that the scatter derived for the small GRB-SN \& SN IcBL sample modelled here can be thought to be representative of its entire population.  This result has encouraging potential for their future use to constrain cosmological models over a larger redshift range to the same degree of precision as SNe Ia.

\section{Conclusions \& Future Research}
\label{sec:Conclusions}

A summary of our results are as follows:

\begin{itemize}
 \item We searched for luminosity--decline relationships for a sample of $N=38$ hydrogen-deficient SNe (six GRB-SNe, 11 SNe Ib, five SNe Ic, eight SNe IIb, four SLSNe-Ic and four SNe IcBL).  Statistically significant relations were seen in filters $BVR$ for the combined relativistic SN IcBL sample, $B$ and $V$ for just the GRB-SN sample, and in the $R$-band of the SLSN-Ic sample, at the $p=0.02$ confidence level.  No statistically significant relations were seen for any of the other SN subtypes.
 \item Two different fitting techniques were used in our analysis: a MC--LLS method and a Bayesian inference method (using MultiNest in the \textsc{python} code environment via PyMultiNest).  Both techniques give the same results when applied to large datasets, as done for the $N=318$ SN Ia dataset from Betoule et al. (2014).  However, the Bayesian inference approach produced poorer results when applied to datasets that displayed a lot of scatter.  The poor quality of the fit was reflected in the larger rms values of the Bayesian procedure relative to the MC--LLS Bootstrap method.
 \item Using the sample of relativistic SNe IcBL, we found a weighted-average Hubble constant of ${H}_{0,\rm w}=82.5\pm8.1$~km~s$^{-1}$~Mpc$^{-1}$.  This value is just over 1$\sigma$ larger than that determined using SNe Ia ($H_{0}=73.8\pm2.4$~km~s$^{-1}$~Mpc$^{-1}$, Riess et al. 2011), and almost 2$\sigma$ larger than that determined by Planck ($H_{0}=67.3$~km~s$^{-1}$~Mpc$^{-1}$) and a combined baryon acoustic oscillations (BAO) + CMB + SN Ia analysis ($H_{0}=69.6\pm0.7$~km~s$^{-1}$~Mpc$^{-1}$; Bennett et al. 2014).
 \item The larger value of $H_{0}$ found here can be attributed to the fact that the host galaxies of relativistic type IcBL SN~2009bb and SN~2012ap are members of galaxy groups, where gravitational interactions with/within their respective group imparts them with non-negligible peculiar motions, and are thus not good indicators of the velocity of the local Hubble flow.
 \item The error in $H_{0}$ determined using SNe IcBL can be reduced by (1) smaller uncertainties in the host galaxy distance measurements, which dominate the error calculation, and (2) larger sample sizes, which will reduce the error in the determined offset of the Hubble ridge line.
 \item The rms scatter in the value of $H_{0}$ determined using SNe Ib, Ic and IIb is of order $20-40$~km~s$^{-1}$~Mpc$^{-1}$, which highlights their unsuitability to be used as standard(izable) candles.
 
\end{itemize}

Our pilot study has demonstrated the suitability of using relativistic SNe IcBL as cosmological probes, in the same fashion as SNe Ia, to determine the value of $H_{0}$ in the local universe.  Indeed our method is essentially identical to those used in historical and current analyses of SNe Ia.  The value of $H_{0}$ found here has an uncertainty of $\sim10$\%, which is greater than determined using combined methods such as BAO+CMB+SN Ia give uncertainties at the $\sim1$\% level (Bennett et al. 2014), and just SNe Ia at the $\sim3$\% level (Riess et al. 2011).  The error in $H_{0}$ determined with SNe IcBL can be reduced with (1) larger sample sizes, and (2) more accurate distances to their host galaxies.

Purists such as ourselves are still interested in determining any possible uncertainties associated with grouping \emph{all} relativistic SNe IcBL together, and instead focusing on just using GRB-SNe to measure $H_{0}$.  However, in order for this task to be completed, an independent distance measurement to the host galaxy of at least one, if not more, GRB-SN host galaxy is required -- none currently exist however.  The closest GRB, and indeed GRB-SN, to date is GRB~980425 \& SN~1998bw, which has a redshift of $z=0.0085$, and, based on a standard cosmology, is likely to be 35--40~Mpc distant.  To date there are several methods for determining the distances to objects at distances $>$30 Mpc, namely surface-brightness fluctuations, the Tully-Fisher relation for spiral galaxies and long-period Cepheid variable stars, the latter of which can in theory be used for distances up to 100~Mpc (Bird et al. 2009).  At~$\approx$~40~Mpc, detecting and monitoring Cepheid variable stars is likely to be quite challenging if attempted with current technology, and is likely only possible to perform using the Hubble Space Telescope (HST), though such a campaign will require 40 or more orbits, which is observationally expensive.  However, HST's successor, the James Webb Space Telescope (JWST), will be much better suited for this task.  Indeed, the NIR and IR capabilities of JWST will be perfectly suited for observing Cepheid variables at $IJK$ wavelengths, where the Period--Luminosity and Period--Wesenheit relations have been accurately determined (e.g. Freedman et al. 2001; Inno et al. 2014), albeit with smaller amplitudes in brightness than bluer optical filters, but which also include corrections for metallicity gradients within a host galaxy (e.g. Zaritsky et al. 1994; Scowcroft et al. 2009).  Moreover, the observations can be obtained at a fraction of the observational cost relative to HST.

Moreover, it is expected that JWST has the potential to observe GRB-SNe up to redshifts\footnote{Determined using the JWST exposure time calculator at \url{http://jwstetc.stsci.edu/etc/.}} of $z=$~3--5.  The only limiting factor now is how to determine independent distances to each high-$z$ object.  One method would be to use the expanding photosphere method (EPM; Kirshner \& Kwan 1974; Schmidt, Kirshner \& Eastman 1992; Schmidt et al. 1994; Eastman, Schmidt \& Kirshner 1996; Vink\'o et al. 2004; Dessart \& Hillier 2005), where one can apply a kinematic model to relate the change in radius of the expanding photosphere to an absolute distance, providing accurate radial velocity measurements are available.  The EPM method has been used extensively for SNe II (e.g. Branch et al. 1981; Schmidt et al. 1994; Hamuy et al. 2001; Leonard et al. 2002), and can achieve distances to accuracies of $\sim$20\% (Hamuy 2001).  The EPM has not been used for stripped-envelope SNe except for SN~2002ap by Vink\'o et al. (2004), who were forced to make the (incorrect) assumption of purely blackbody emission in order to derive a distance to its host galaxy M74.  The success of using the EPM for SNe II is based on calculations made of the dilution factor, which is sometimes referred to as the distance-correction factor, which for SNe II has been determined via radiative-transfer (RT) simulations (e.g. Eastman, Schmidt \& Kirshner 1996; Dessart \& Hillier 2005).  The dilution factor is necessary in order to account for the amount of flux dilution that occurs in SN atmospheres that are dominated by electron scattering.  The physics underlying the dilution factor is complex, and its precise value depends on the temperature, composition and density of the SN atmosphere, as well as the thermalization radius.   The EPM has a lot of potential for GRB-SNe however, for all one needs are well sampled LCs in at least two filters (preferably more to construct a bolometric LC), a time-series spectra to estimate the photospheric velocity as a function of time, and accurately determined values of the dilution factor for a wide range of SN atmospheres.  The necessary observations will be obtainable with JWST at the end of the decade, therefore in the meantime advances in RT simulations are needed to determine the dilution factor for SNe Ibc over a wide range of SN compositions and arrangements (which has been partially calculated by Dessart et al. 2015; but these are not suitable for application to relativistic SNe IcBL), especially for GRB-SNe.  Should such advances be made, then GRB-SNe have the potential to be used over a wider redshift range than SNe Ia.  Moreover, the precise localization of GRBs by dedicated GRB satellites means that finding high-$z$ GRB-SN can be achieved with a moderate about of ease (along with a healthy dose of patience waiting for a high-$z$ event to occur), especially when compared with the very hard task, and ambiguities associated with finding SNe Ia at the same (large) redshifts.

One final musing needs to be contemplated when deciding whether to continue using relativistic SNe IcBL as cosmological probes.  The physical conditions that produce these energetic SNe may be rather regular -- the distribution of peak brightnesses and bolometric properties are reminiscent of those of SNe Ia, which themselves are thought to arise from somewhat ``regular'' physical conditions.  However, recent observations suggest that LGRBs and hence GRB-SNe arise in turbulent conditions (e.g. Kelly et al. 2014), although not as turbulent as SLSNe (e.g. Leloudas et al. 2015).  In many situations, this turbulence may be the consequence of interacting galaxies.  Indeed it has been shown that many GRB host galaxies are part of interacting systems, with up to a third or more showing signs of galactic interaction (e.g. Wainwright et al. 2007; Kelly et al. 2008).  Such interactions will clearly impart peculiar velocities to the galaxy itself, putting it at odds to the local Hubble flow, thus limiting the usefulness the SN harboured within as a cosmological candle.  However, at larger distances from Earth, the peculiar velocities become increasingly less important, and distances greater than 100~Mpc or so, are already less then 10\% of the velocity due to cosmological expansion\footnote{For $H_{0} = 70$~km~s$^{-1}$~Mpc$^{-1}$, the recessional velocity of a galaxy at 100~Mpc is 7000~km~s$^{-1}$.  Taking the peculiar velocity of the Local Group as an example (630~km~s$^{-1}$, towards the Virgo cluster), at 100~Mpc this is less than 10\% of the velocity of cosmological expansion.}.  Therefore, methods for determining the distances to more distant objects, perhaps those based on the EPM (though keeping in mind the limitations of such approaches; e.g. Dessart \& Hillier 2005; Dessart et al. 2015) or perhaps Cepheids using JWST, can ultimately prove the usefulness of GRB-SNe, and indeed all relativistic SNe IcBL, as cosmological probes.

\section{Acknowledgements}

We thank Gulli Bj\"ornsson for his very helpful comments to the original manuscript.  ZC and PJ gratefully acknowledge support from a Project Grant from the Icelandic Research Fund.

\appendix

\section{Fitting Methods}
\label{sec:Appendix_methods}

\subsection{Monte-Carlo Linear-Least Squares Bootstrap Method}

The first fitting method we utilized was a bootstrap algorithm that used Monte-Carlo (MC) sampling and replacement along with equal-weighted linear regression data-fitting, and was written using \textsc{python}.  In this approach, a new dataset is created from the original by choosing a new value from a Gaussian that is centered on each original datapoint, and whose standard deviation in each direction is equal to the error in that variable.  The simulation was performed 10,000 times, where each time a straight line was fit to the new dataset using an equally weighted LLS optimized curve algorithm (\textsc{scipy.optimize.curve$\_$fit}); i.e. the Levenberg-Marquardt method (Levenberg 1944), which is a standard technique for solving nonlinear least squares problems.  We used the standard deviation of the distribution of values as an estimate of the standard (1$\sigma$, 68\% confidence interval) error of each free parameter.

\subsection{Bayesian--MultiNest Method}

The second fitting method we employed was a Bayesian inference procedure based on the MultiNest algorithm created by Feroz et al. (2009), and made usable in \textsc{python} via the PyMultiNest package (Buchner et al. 2014).  The MultiNest algorithm is based on the idea of nested sampling (Skilling 2004), which is a method targeted at the efficient calculation of the Bayesian evidence (i.e. the average likelihood of a model/parameter value over its prior probability space; thus evidence can be used to assign relative probabilities to different models/parameters), which produces posterior inferences as a by-product.  Inferences are obtained by taking samples from the unnormlised posterior using standard Markov-Chain MC methods, where at equilibrium the chain contains a set of samples from the parameter space that is distributed according to the posterior.  This posterior constitutes the complete Bayesian inference of the parameter values, and can be marginalized over each parameter to obtain individual parameter constraints.  The nested sampling algorithm uses an elliptical bound that contains the current point set at each stage of the fitting process, where in each stage the region is restricted to smaller regions around the posterior peak from which new samples are drawn.  The MultiNest algorithm expands this method to a multi-nodal nested sampling algorithm that calculates the evidence as well as an associated error estimate.  For further details we refer the reader to the aforementioned papers that describe in detail the nuances of the nested sampling algorithm.

When performing the various fits for the luminosity--decline relationship and the intercept of the Hubble ridge line, we also performed a sensitivity analyse where we chose several different prior ranges to see if larger or smaller ranges affected the final value of the fitted parameters.  We found that our result was completely independent of the prior range used.  Note also that the errors derived by the PyMultiNest are 1$\sigma$ (68\% confidence interval) errors.

\section{Hydrogen-deficient supernovae}
\label{sec:Appendix_Ibc}

\subsection{Luminosity--decline relationships}
\label{sec:Appendix_lum_decline_Ibc}

\begin{figure*}
 \centering
 \includegraphics[bb=0 0 880 1001, scale=0.53]{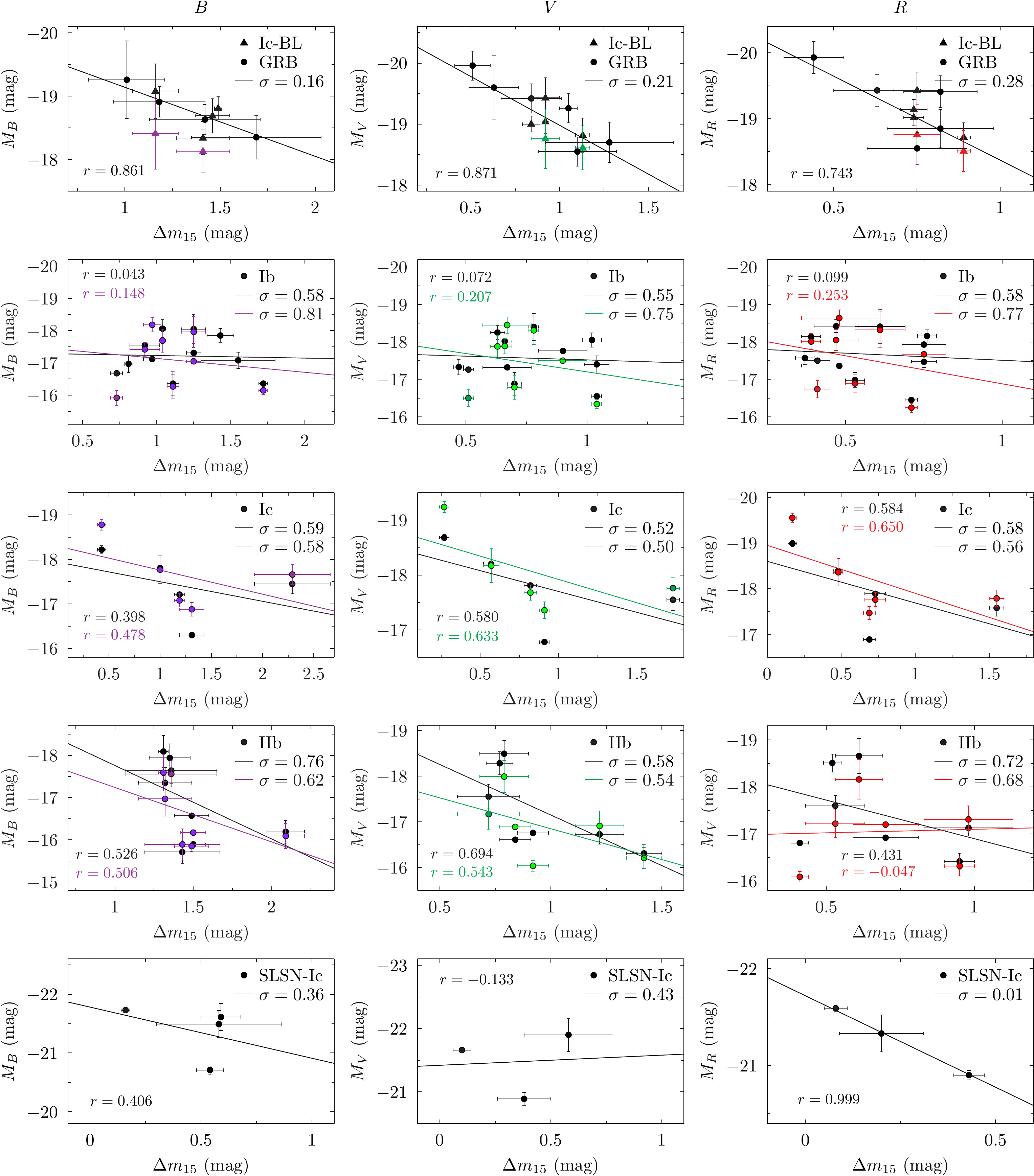}
 \caption{Luminosity--decline relationships for SNe Ib, Ic, IIb, SLSN-Ic, IcBL and GRB-SNe in filters $BVR$ (purple, green, red, respectively).  As in Fig. \ref{fig:lum_decline_IcBL}, the solid black lines and points correspond to absolute magnitudes calculated for luminosity distances (see Section \ref{sec:linear_distances}), while coloured points and lines correspond to absolute magnitudes calculated for those events where an independent distance measurement(s) have been made to the SN's host galaxy.  The correlation coefficient ($r$) for each dataset is shown (in black and in their respective colours) as well as the best-fitting luminosity--decline relationship (equ. \ref{equ:peak_m15}) determined using the MC--LLS bootstrap method, and the corresponding rms ($\sigma$) of the fitted model.  It is seen that the strongest correlations are present for the GRB-SNe and combined GRB-SN \& SN IcBL samples (which can be considered to be statistically significant -- see Table \ref{table:lum_decline} for their corresponding $p$ values), while a strong correlation is also seen for the SLSNe-Ic in the $R$-band. }
 \label{fig:lum_decline_all}
\end{figure*}

Figure \ref{fig:lum_decline_all} shows SNe Ib, Ic, IIb, IcBL, GRB-SNe and SLSNe-Ic in the $BVR$ absolute magnitude--$\Delta m_{15}$ parameter space.  A statistically significant correlation is only seen for the GRB-SNe, SNe IcBL (and combined sample), and SLSNe-Ic in the $R$-band only.  The correlation coefficient ($r$) for each dataset is shown (in black and in their respective colours) as well as the best-fitting luminosity--decline relationship (equ. \ref{equ:peak_m15}) determined using the MC--LLS bootstrap method, and the corresponding rms ($\sigma$) of the fitted model.  All fitted parameters can be found in Table \ref{table:lum_decline}.

Inserra \& Smartt (2014) demonstrated that SLSNe-Ic have a luminosity--decline relationship.  Using a sample of 16 SLSNe-Ic they showed a relationship between the peak absolute magnitude at 400~nm (between the Johnsons $U$ and $B$ filters) and different decay rates (10, 20, and 30 days from peak; compared with 15 days considered here).  As considered here, they also computed K-corrections before searching for relationships, and they considered SNe regardless of redshift.  In this work, we have maintained an inclusion in our sample of $z\le0.2$, which excludes the vast majority of SLSNe-Ic in their sample.  As such, we have only investigated four events, where only three of them have observations in all three $BVR$ filters.  

In the $B$-band, there is a hint of a correlation present between $M_{B}$ and $\Delta m_{15,B}$, and certainly a clear correlation is seen in the $R$-band.  However, we have limited ourselves to any conclusions that can be drawn from investigating such a small dataset, and the results of Inserra \& Smartt (2014) are not refuted in this paper.  Certainly our correlation in the red $R$-band filter appears to concur with their results, but we loathe to drawn such a comparison based on only three datapoints.  Finally, we also hesitate to draw any firm conclusions due to the unknown host-extinction of many of the SLSNe used in their analysis (which is clearly pointed out by the authors as well), which is a key requirement for any SN to be included in our analysis.

\subsection{Hubble diagrams}
\label{sec:Appendix_Hubble_Ibc}

The weighted and average values of $H_{0}$ obtained from the SN Ib sample determined using the MC--LLS bootstrap method ($H_{0,\rm w}=70.6\pm1.0$~km~s$^{-1}$~Mpc$^{-1}$, and $\bar{H}_{0}=78.9$~km~s$^{-1}$~Mpc$^{-1}$, with a standard deviation of 31.3 km~s$^{-1}$~Mpc$^{-1}$) are consistent with that determined using SNe Ia ($H_{0}=73.8\pm2.4$~km~s$^{-1}$~Mpc$^{-1}$; Riess et al. 2011).  However, when using the Bayesian--MultiNest approach, the average and weighted values of $H_{0}$ is larger.  This is attributed to the fact that the Bayesian approach does not perform as well as the MC bootstrap method when fitting straight lines to data that are not well correlated and have a lot of scatter.  This happens regardless of the range of priors used in the fitting algorithm.  The poorer performance of the Bayesian method is reflected in the rms values of the fitted line determined using PyMultiNest, which are larger than those determined using the MC--LLS bootstrap method (see for example, the SNe Ic sample in Tables \ref{table:lum_decline}, where the values of $\sigma$ are much larger for the Bayesian method). 

For the SN Ic samples, the weighted values of $H_{0}$ are smaller than those determine using the other datasets ($H_{0,\rm w}\approx50$~km~s$^{-1}$~Mpc$^{-1}$), although the average values are closer to those of the other SNe.  Poor agreement is seen between the MC--LLS bootstrap and Bayesian approaches, as was seen for the SN Ib sample.  In contrast, the SN IIb sample finds a much larger value for the Hubble constant, reaching a value of $H_{0}\approx80$~km~s$^{-1}$~Mpc$^{-1}$ in the MC--LLS bootstrap approach, and exceeding 120~km~s$^{-1}$~Mpc$^{-1}$ in the Bayesian approach.  The average values of $H_{0}$ determined using the two SNe IcBL is also higher than the generally accepted values, where we find average and weighted average values of $H_{0}\approx80$~km~s$^{-1}$~Mpc$^{-1}$.  Clearly, the SNe Ib produce a value of $H_{0}$ that is in closest agreement with values determined via other methods (e.g. the cosmic microwave background, SNe Ia, etc.)

\begin{figure*}
 \centering
 \includegraphics[bb=0 0 779 1561, scale=0.41]{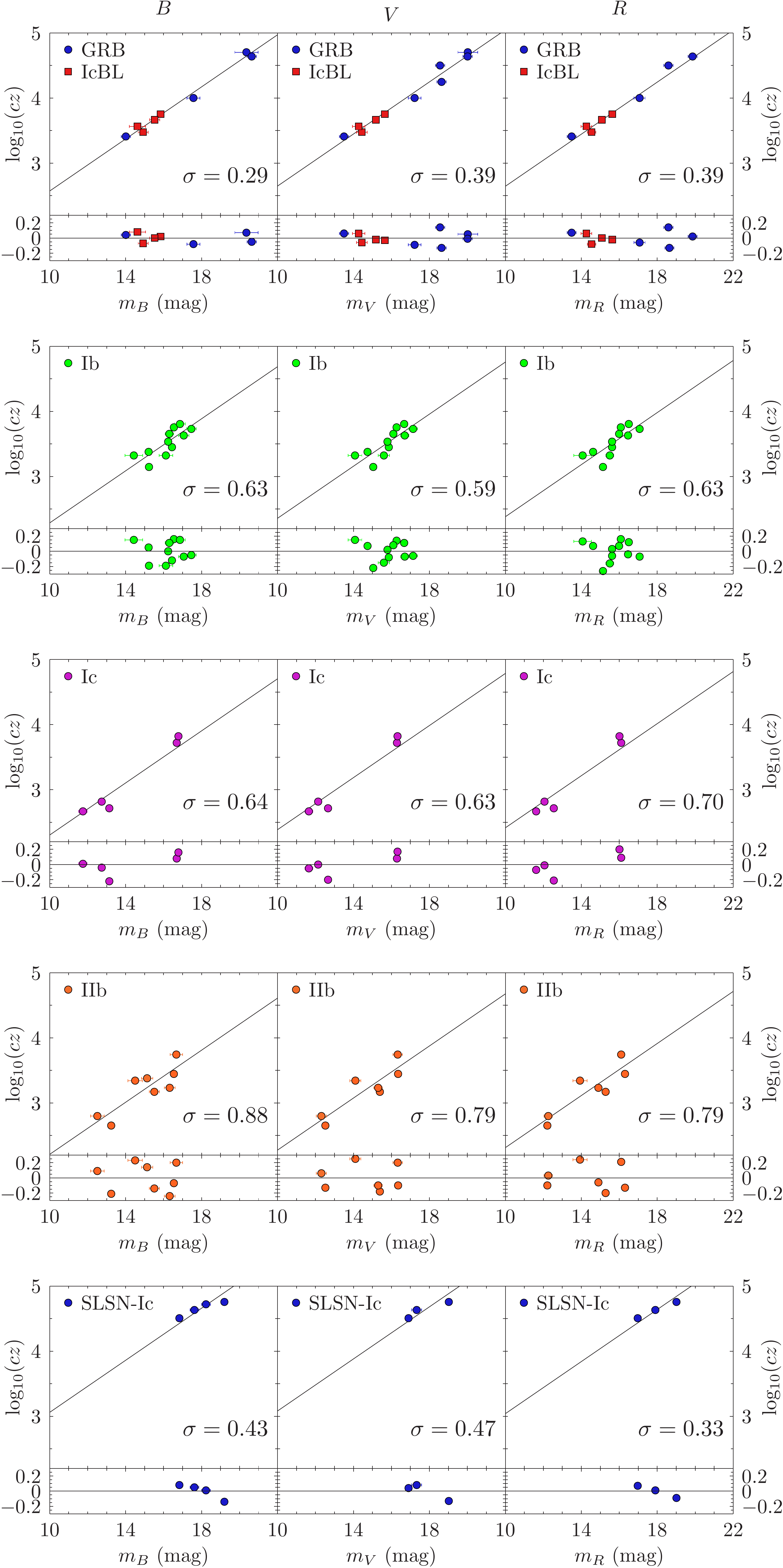}
 \caption{Hubble diagrams of hydrogen-deficient SNe.  Plotted in each subplot are the uncorrected magnitudes of each subtype and the fitted Hubble ridge line as determined using the MC--LLS bootstrap method.  Also plotted are the rms values and residuals of the magnitudes about the ridge line.  It is seen that the combined GRB-SN \& SN IcBL sample has the smallest scatter, while the SLSN-Ic sample as exhibits lower scatter than the SNe Ib, Ic and IIb samples.  }
 \label{fig:hubble_diagrams}
\end{figure*}

\subsection{The curious case of SN 1994I}
\label{sec:Appendix_1994I}

\begin{figure*}
 \centering
 \includegraphics[bb=0 0 878 398, scale=0.52]{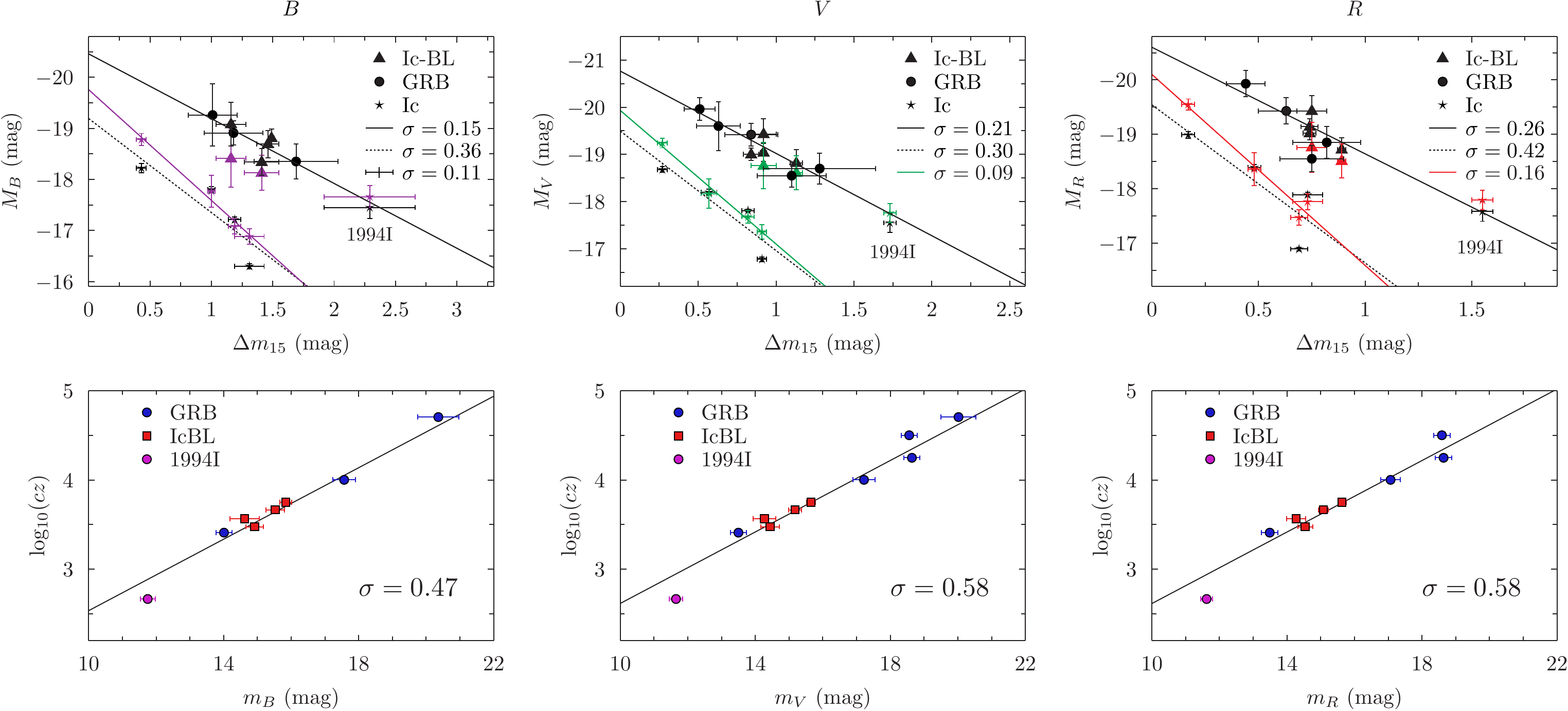}
 \caption{The curious case of type Ic SN 1994I.  \textbf{Top panels:}  SN~1994I appears to follow the same luminosity--decline relationship as GRB-SNe and SNe IcBL -- indeed statistically significant relations (at the $p=0.02$ significance level) that include SN~1994I in the GRB-SNe and SNe IcBL samples are seen in all three filters. Moreover, the luminosity--decline relationships of the SN Ic samples that exclude SN~1994I are tighter and exhibit less scatter, but they are not statistically significant.  \textbf{Bottom panels:}  However, including SN~1994I in the GRB-SNe and SNe IcBL samples in the Hubble diagrams creates much larger rms values that are roughly a factor of 1.5 larger than the combined GRB-SN \& SN IcBL samples that do not include SN~1994I.  We therefore conclude that it is only a fluke that SN~1994I obeys the same luminosity--decline relationship as the GRB-SN \& SN IcBL samples, and it does not represent a relativistic explosion.}
 \label{fig:1994I}
\end{figure*}

Close inspection of the SNe Ic in Fig. \ref{fig:lum_decline_all} shows that there is an outlier that has much more rapid decline (larger $\Delta m_{15}$) parameters than the other SNe Ic.  This outlier is SN~1994I, which for many years was thought to be the archetype type Ic SN,, though recently opinion has shifted towards the notion that it is likely anything \textit{but} a ``typical'' type Ic SN (e.g. Modjaz et al. 2015).  

In our investigation we found that SN~1994I appears to obey the same luminosity--decline relationship as the relativistic SN sample (GRB-SNe \& SNe IcBL) -- see top row in Fig. \ref{fig:1994I}.  When SN~1994I is included, statistically significant luminosity--decline relationships are seen in all three filters at a significance level of $p=0.02$ ($r=0.956,0.943,0.903$ in filters $BVR$, respectively).  The scatter of the relationships are almost identical to the combined GRB-SN \& SN IcBL samples ($\sigma=0.15,0.21,0.26$ mag in $BVR$, respectively; compared with their values of $\sigma=0.16,0.21,0.28$ mag in Table \ref{table:lum_decline}). 

Additionally, the luminosity--decline relationships of the SN Ic sample that excludes SN~1994I are tighter and exhibit less scatter than the SNe Ic sample that includes SN~1994I ($\sigma=0.59,0.52,0.58$ mag when SN~1994I is included and $\sigma=0.36,0.30,0.42$ mag when it is omitted).  However, the relationships are not statistically significant ($r=0.869,0.903,0.839$ for $N=4$ objects).

Motivated by the fact that SN~1994I appears to follow the same luminosity--decline relationship as the relativistic SNe, we included it in the Hubble diagram (see bottom of Fig. \ref{fig:1994I}).  Here it is seen that including SN~1994I with the combined GRB-SN \& SN IcBL sample, the rms scatter increases by almost a factor of 1.5 ($\sigma=0.47,0.58,0.58$ mag in filters $BVR$ when it was included, and $\sigma=0.29,0.39,0.38$ mag when it was omitted).  When SN~1994I was omitted from the SN Ic sample on the Hubble diagram (not plotted), the scatter of the sample about the Hubble ridge line actually increased ($\sigma=0.64,0.63,0.70$ mag in filters $BVR$, respectively, when it was included, and $\sigma=0.71,0.69,0.76$ mag, in filters $BVR$ respectively, when it was omitted).  We therefore conclude that it is merely a fluke that SN~1994I obeys the same luminosity--decline relationship as the combined GRB-SN \& SN IcBL sample, and it does not represent a relativistic SN.

\section{Data Tables}

\begin{landscape}

\begin{table}
\small
\centering
\caption{Photometric Properties of SNe Ib, Ic, IIb, SLSN-Ic, IcBL and GRB-SNe}
\setlength{\tabcolsep}{4.5pt}
\label{table:photometry}
\begin{tabular}{|cccccccccccc|}
\hline
 &  &  & (mag) & (mag) & (mag) & (mag) & (mag) & (mag) & (mag) & (mag) & \\
SN	&	Type	&		$z$				&	$E(B-V)_{\rm fore}$	&		$E(B-V)_{\rm rest}$				&		$m_{B}$ (mag)				&		$m_{V}$				&		$m_{R}$ 			&		$\Delta m_{15,B}$ 			&		$\Delta m_{15,V}$ 			&		$\Delta m_{15,R}$ 			&	Ref(s).	\\
\hline
1999dn	&	Ib	&	$	0.00938	\pm	0.0001	$	&	0.05	&	$	0.05			$	&	$	16.44	\pm	0.05	$	&	$	15.86	\pm	0.04	$	&	$	15.62	\pm	0.06	$	&	$	0.73	\pm	0.04	$	&	$	0.51	\pm	0.02	$	&	$	0.41	\pm	0.04	$	&	(1--2)	\\
1999ex	&	Ib	&	$	0.011401	\pm	0.0001	$	&	0.02	&	$	0.28			$	&	$	16.24	\pm	0.03	$	&	$	15.79	\pm	0.03	$	&	$	15.62	\pm	0.02	$	&	$	1.25	\pm	0.05	$	&	$	0.90	\pm	0.10	$	&	$	0.75	\pm	0.07	$	&	(3)	\\
2005bf$^{*}$	&	Ib	&	$	0.018913	\pm	0.0001	$	&	0.04	&	$	0.00			$	&	$	16.54	\pm	0.04	$	&	$	16.27	\pm	0.02	$	&	$	16.08	\pm	0.02	$	&	$	0.97	\pm	0.06	$	&	$	0.67	\pm	0.10	$	&	$	0.48	\pm	0.12	$	&	(4)	\\
2005hg	&	Ib	&	$	0.0213	\pm	0.0001	$	&	0.09	&	$	0.25	\pm	0.07^{\dagger}	$	&	$	16.86	\pm	0.28	$	&	$	16.67	\pm	0.19	$	&	$	16.50	\pm	0.17	$	&	$	1.04	\pm	0.02	$	&	$	0.63	\pm	0.03	$	&	$	0.47	\pm	0.07	$	&	(5)	\\
2006ep	&	Ib	&	$	0.0150	\pm	0.0001	$	&	0.04	&	$	0.45	\pm	0.07^{\dagger}	$	&	$	16.30	\pm	0.22	$	&	$	16.10	\pm	0.19	$	&	$	15.99	\pm	0.16	$	&	$	1.43	\pm	0.09	$	&	$	1.02	\pm	0.04	$	&	$	0.76	\pm	0.03	$	&	(5)	\\
2007kj	&	Ib	&	$	0.0179	\pm	0.0001	$	&	0.07	&	$	0.10	\pm	0.07^{\dagger}	$	&	$	17.46	\pm	0.25	$	&	$	17.14	\pm	0.23	$	&	$	17.07	\pm	0.15	$	&	$	1.55	\pm	0.25	$	&	$	1.04	\pm	0.05	$	&	$	0.75	\pm	0.04	$	&	(5)	\\
2007uy	&	Ib	&	$	0.0070	\pm	0.0001	$	&	0.02	&	$	0.60	\pm	0.13	$	&	$	14.43	\pm	0.46	$	&	$	14.08	\pm	0.36	$	&	$	14.07	\pm	0.46	$	&	$	1.25	\pm	0.08	$	&	$	0.78	\pm	0.02	$	&	$	0.61	\pm	0.08	$	&	(5)	\\
2007Y	&	Ib	&	$	0.004657	\pm	0.0001	$	&	0.02	&	$	0.09			$	&	$	15.23	\pm	0.04	$	&	$	15.04	\pm	0.03	$	&	$	15.14	\pm	0.02	$	&	$	1.72	\pm	0.03	$	&	$	1.04	\pm	0.02	$	&	$	0.71	\pm	0.02	$	&	(6)	\\
2008D	&	Ib	&	$	0.0070	\pm	0.0001	$	&	0.02	&	$	0.60	\pm	0.10	$	&	$	16.12	\pm	0.36	$	&	$	15.60	\pm	0.30	$	&	$	15.50	\pm	0.20	$	&	$	1.11	\pm	0.04	$	&	$	0.70	\pm	0.03	$	&	$	0.53	\pm	0.03	$	&	(7)	\\
2009iz	&	Ib	&	$	0.01419	\pm	0.0001	$	&	0.07	&	$	0.03	\pm	0.07^{\dagger}	$	&	$	17.06	\pm	0.27	$	&	$	16.70	\pm	0.21	$	&	$	16.46	\pm	0.17	$	&	$	0.81	\pm	0.03	$	&	$	0.47	\pm	0.02	$	&	$	0.37	\pm	0.03	$	&	(5)	\\
2009jf	&	Ib	&	$	0.007942	\pm	0.0001	$	&	0.10	&	$	0.00			$	&	$	15.21	\pm	0.03	$	&	$	14.74	\pm	0.02	$	&	$	14.62	\pm	0.02	$	&	$	0.92	\pm	0.10	$	&	$	0.66	\pm	0.03	$	&	$	0.39	\pm	0.03	$	&	(8)	\\
1994I	&	Ic	&	$	0.00155	\pm	0.0001	$	&	0.03	&	$	0.42			$	&	$	11.75	\pm	0.22	$	&	$	11.65	\pm	0.20	$	&	$	11.62	\pm	0.18	$	&	$	2.29	\pm	0.37	$	&	$	1.73	\pm	0.04	$	&	$	1.55	\pm	0.05	$	&	(9)	\\
2002ap	&	Ic	&	$	0.002187	\pm	0.0001	$	&	0.06	&	$	0.03			$	&	$	12.74	\pm	0.05	$	&	$	12.14	\pm	0.03	$	&	$	12.06	\pm	0.04	$	&	$	1.19	\pm	0.05	$	&	$	0.82	\pm	0.04	$	&	$	0.73	\pm	0.07	$	&	(10--13)	\\
2004aw	&	Ic	&	$	0.0175	\pm	0.0001	$	&	0.02	&	$	0.00			$	&	$	16.69	\pm	0.06	$	&	$	16.29	\pm	0.06	$	&	$	16.10	\pm	0.05	$	&	$	1.00	\pm	0.03	$	&	$	0.57	\pm	0.05	$	&	$	0.48	\pm	0.03	$	&	(14)	\\
2007gr	&	Ic	&	$	0.001729	\pm	0.0001	$	&	0.05	&	$	0.03			$	&	$	13.14	\pm	0.05	$	&	$	12.66	\pm	0.05	$	&	$	12.55	\pm	0.03	$	&	$	1.31	\pm	0.12	$	&	$	0.91	\pm	0.03	$	&	$	0.69	\pm	0.04	$	&	(15--16)	\\
2011bm	&	Ic	&	$	0.0221	\pm	0.0001	$	&	0.03	&	$	0.03			$	&	$	16.78	\pm	0.09	$	&	$	16.32	\pm	0.06	$	&	$	16.01	\pm	0.06	$	&	$	0.43	\pm	0.04	$	&	$	0.27	\pm	0.03	$	&	$	0.17	\pm	0.03	$	&	(17)	\\
2006T	&	IIB	&	$	0.0080	\pm	0.0001	$	&	0.06	&	$	0.37	\pm	0.07^{\dagger}	$	&	$	15.13	\pm	0.30	$	&	$	-			$	&	$	-			$	&	$	1.36	\pm	0.29	$	&	$	-			$	&	$	-			$	&	(5)	\\
2008ax	&	IIB	&	$	0.0021	\pm	0.0001	$	&	0.02	&	$	0.48	\pm	0.10	$	&	$	12.51	\pm	0.36	$	&	$	12.31	\pm	0.27	$	&	$	12.26	\pm	0.22	$	&	$	1.32	\pm	0.17	$	&	$	0.72	\pm	0.14	$	&	$	0.53	\pm	0.10	$	&	(18--19)	\\
2008bo	&	IIB	&	$	0.0049	\pm	0.0001	$	&	0.05	&	$	0.22	\pm	0.07^{\dagger}	$	&	$	15.51	\pm	0.27	$	&	$	15.39	\pm	0.19	$	&	$	15.28	\pm	0.17	$	&	$	2.09	\pm	0.12	$	&	$	1.42	\pm	0.08	$	&	$	0.95	\pm	0.05	$	&	(5)	\\
2010as	&	IIB	&	$	0.007354	\pm	0.0001	$	&	0.15	&	$	0.42	\pm	0.10	$	&	$	14.50	\pm	0.38	$	&	$	14.10	\pm	0.29	$	&	$	13.93	\pm	0.37	$	&	$	1.31	\pm	0.03	$	&	$	0.79	\pm	0.11	$	&	$	0.61	\pm	0.08	$	&	(20)	\\
2011dh	&	IIB	&	$	0.00155	\pm	0.0001	$	&	0.03	&	$	0.00			$	&	$	13.24	\pm	0.02	$	&	$	12.52	\pm	0.05	$	&	$	12.21	\pm	0.02	$	&	$	1.50	\pm	0.08	$	&	$	0.84	\pm	0.07	$	&	$	0.70	\pm	0.11	$	&	(21)	\\
2011ei	&	IIB	&	$	0.009317	\pm	0.0001	$	&	0.05	&	$	0.18			$	&	$	16.54	\pm	0.04	$	&	$	16.35	\pm	0.04	$	&	$	16.30	\pm	0.03	$	&	$	1.49	\pm	0.11	$	&	$	0.92	\pm	0.07	$	&	$	0.41	\pm	0.03	$	&	(22)	\\
2011fu	&	IIB	&	$	0.018489	\pm	0.000037	$	&	0.06	&	$	0.16	\pm	0.11	$	&	$	16.67	\pm	0.33	$	&	$	16.33	\pm	0.25	$	&	$	16.10	\pm	0.19	$	&	$	1.35	\pm	0.13	$	&	$	0.77	\pm	0.06	$	&	$	0.52	\pm	0.03	$	&	(23)	\\
2011hs	&	IIB	&	$	0.005701	\pm	0.0001	$	&	0.01	&	$	0.16	\pm	0.08	$	&	$	16.32	\pm	0.28	$	&	$	15.30	\pm	0.22	$	&	$	14.90	\pm	0.17	$	&	$	1.43	\pm	0.24	$	&	$	1.22	\pm	0.11	$	&	$	0.98	\pm	0.15	$	&	(24)	\\
PTF11rks	&	SLSN-Ic	&	$	0.19	\pm	0.0001	$	&	0.01	&	$	-			$	&	$	16.83	\pm	0.02	$	&	$	16.90	\pm	0.03	$	&	$	16.97	\pm	0.03	$	&	$	0.16	\pm	0.02	$	&	$	0.10	\pm	0.04	$	&	$	0.08	\pm	0.03	$	&	(25)	\\
2011ke	&	SLSN-Ic	&	$	0.143	\pm	0.0001	$	&	0.01	&	$	-			$	&	$	17.62	\pm	0.23	$	&	$	17.33	\pm	0.26	$	&	$	17.90	\pm	0.19	$	&	$	0.59	\pm	0.09	$	&	$	0.58	\pm	0.20	$	&	$	0.20	\pm	0.11	$	&	(25)	\\
2012il	&	SLSN-Ic	&	$	0.175	\pm	0.0001	$	&	0.02	&	$	-			$	&	$	18.23	\pm	0.23	$	&	$	-			$	&	$	-			$	&	$	0.58	\pm	0.28	$	&	$	-			$	&	$	-			$	&	(25)	\\
PTF12dam	&	SLSN-Ic	&	$	0.107	\pm	0.0001	$	&	0.04	&	$	-			$	&	$	19.20	\pm	0.07	$	&	$	19.02	\pm	0.10	$	&	$	19.01	\pm	0.05	$	&	$	0.54	\pm	0.06	$	&	$	0.38	\pm	0.12	$	&	$	0.43	\pm	0.04	$	&	(26)	\\
2003jd	&	IcBL	&	$	0.018826	\pm	0.0001	$	&	0.08	&	$	0.06	\pm	0.10	$	&	$	15.84	\pm	0.18	$	&	$	15.65	\pm	0.13	$	&	$	15.63	\pm	0.12	$	&	$	1.49	\pm	0.04	$	&	$	0.84	\pm	0.05	$	&	$	0.74	\pm	0.03	$	&	(27)	\\
2007ru	&	IcBL	&	$	0.01546	\pm	0.0001	$	&	0.23	&	$	0.04	\pm	0.06	$	&	$	15.53	\pm	0.27	$	&	$	15.18	\pm	0.19	$	&	$	15.08	\pm	0.15	$	&	$	1.46	\pm	0.09	$	&	$	0.92	\pm	0.02	$	&	$	0.74	\pm	0.04	$	&	(28)	\\
2009bb	&	IcBL	&	$	0.009987	\pm	0.0001	$	&	0.08	&	$	0.50	\pm	0.07	$	&	$	14.92	\pm	0.26	$	&	$	14.44	\pm	0.28	$	&	$	14.54	\pm	0.22	$	&	$	1.41	\pm	0.14	$	&	$	1.13	\pm	0.04	$	&	$	0.89	\pm	0.02	$	&	(29--30)	\\
2012ap	&	IcBL	&	$	0.012241	\pm	0.0001	$	&	0.04	&	$	0.83	\pm	0.12	$	&	$	14.62	\pm	0.43	$	&	$	14.27	\pm	0.33	$	&	$	14.27	\pm	0.28	$	&	$	1.16	\pm	0.12	$	&	$	0.92	\pm	0.08	$	&	$	0.75	\pm	0.07	$	&	(31--32)	\\
1998bw	&	GRB	&	$	0.00857	\pm	0.00024	$	&	0.05	&	$	variable			$	&	$	14.01	\pm	0.24	$	&	$	13.50	\pm	0.24	$	&	$	13.49	\pm	0.24	$	&	$	1.18	\pm	0.24	$	&	$	0.84	\pm	0.17	$	&	$	0.63	\pm	0.13	$	&	(33)	\\
2003dh	&	GRB	&	$	0.1685	\pm	0.0002	$	&	0.02	&	$	0.14	\pm	0.05	$	&	$	20.36	\pm	0.61	$	&	$	20.02	\pm	0.52	$	&	$	-			$	&	$	1.01	\pm	0.20	$	&	$	0.63	\pm	0.14	$	&	$	-			$	&	(34)	\\
2003lw	&	GRB	&	$	0.1055	\pm	0.00007	$	&	0.90	&	$	0.24	\pm	0.10	$	&	$	-			$	&	$	18.56	\pm	0.24	$	&	$	18.59	\pm	0.24	$	&	$	-			$	&	$	0.51	\pm	0.10	$	&	$	0.44	\pm	0.09	$	&	(35--36)	\\
2006aj	&	GRB	&	$	0.03342	\pm	0.00002	$	&	0.13	&	$	0.05	\pm	0.01	$	&	$	17.57	\pm	0.34	$	&	$	17.22	\pm	0.33	$	&	$	17.07	\pm	0.29	$	&	$	1.69	\pm	0.34	$	&	$	1.28	\pm	0.36	$	&	$	0.82	\pm	0.16	$	&	(37--38)	\\
2010bh	&	GRB	&	$	0.059	\pm	0.0001	$	&	0.10	&	$	0.16	\pm	0.01	$	&	$	-			$	&	$	18.64	\pm	0.24	$	&	$	18.64	\pm	0.24	$	&	$	-			$	&	$	1.10	\pm	0.22	$	&	$	0.75	\pm	0.15	$	&	(39)	\\
2013dx	&	GRB	&	$	0.145	\pm	0.001	$	&	0.04	&	$	0.00			$	&	$	20.64	\pm	0.24	$	&	$	20.01	\pm	0.24	$	&	$	19.86	\pm	0.24	$	&	$	1.42	\pm	0.29	$	&	$	1.05	\pm	0.05	$	&	$	0.82	\pm	0.11	$	&	(40)	\\
\hline
\end{tabular}
\begin{flushleft}
$^{*}$Photometric properties of the second peak; $^{\dagger}$Rest-frame extinction determined using method in Drout et al. (2011). \scriptsize{\textbf{References}: (1) Bennetti et al. (2011); (2)~Cano et al. (2014a); (3) Stritzinger et al. (2002); (4) Folatelli et al. (2006); (5) Bianco et al. (2014); (6) Stritzinger et al. (2009); (7) Malesani et al. (2009); (8)~Valenti et al. (2011); (9) Richmond et al. (1996); (10) Tomita et al. (2006); (11) Yoshii et al. (2003); (12) Foley et al. (2003); (13) Pandey et al. (2003); (14)~Taubenberger et al. (2006); (15) Hunter et al. (2009); (16) Valenti et al. (2008a); (17) Valenti et al. (2012); (18) Taubenberger et al. (2011); (19) Chornock et al. (2011); (20) Folatelli et al. (2014); (21) Sahu et al. (2013); (22) Milisavljevic et al. (2013); (23) Kumar et al. (2013); (24) Bufano et al. (2014); (25) Inserra et al. (2013); (25) Inserra et al. (2013); (25) Inserra et al. (2013); (26) Nicholl et al. (2013); (27) Valenti et al. (2008b); (28) Sahu et al. (2009); (29) Pignata et al. (2011); (30) Levesque et al. (2010); (31) Liu et al. (2015); (32) Milisavljevic et al. (2015); (33) Clocchiatti et al. (2011); (34) Deng et al. (2005); (35) Malesani et al. (2004); (36) Margutti et al. (2007); (37) Sollerman et al. (2006); (38) Ferrero et al. (2006); (39) Olivares E. et al. (2012); (40) D'Elia et al. (2015). }\\
\end{flushleft}
\end{table}

\end{landscape}

\begin{table*}
\small
\centering
\setlength{\tabcolsep}{14pt}
\caption{Linear and Luminosity Distances to SNe Ib, Ic, IIb, SLSN-Ic, IcBL and GRB-SNe}
\label{table:distances}
\begin{tabular}{|ccccccc|}
\hline
SN	&	Type	&	Host	&		$D_{\rm w}$ (Mpc)$^{\dagger}$				&	$D_{L}$ (Mpc)	&	$N_{\rm obs}$	&	Ref(s).	\\
\hline	
1999dn	&	Ib	&	NGC 7714	&	$	29.65	\pm	3.18	$	&	42.08	&	4	&	(1)	\\
1999ex	&	Ib	&	IC 5179	&	$	45.40	\pm	0.25	$	&	51.23	&	25	&	(1-16)	\\
2005bf	&	Ib	&	MCG+00-27-05	&	$	88.10	\pm	9.45	$	&	85.46	&	4	&	(1)	\\
2005hg	&	Ib	&	UGC 1394	&	$	81.29	\pm	8.72	$	&	96.41	&	4	&	(1)	\\
2006ep	&	Ib	&	NGC 214	&	$	-			$	&	67.58	&	--	&	--	\\
2007kj	&	Ib	&	NGC 7803	&	$	-			$	&	80.82	&	--	&	--	\\
2007uy	&	Ib	&	NGC 2770	&	$	30.01	\pm	1.77	$	&	31.35	&	13	&	(1, 15-19)	\\
2007Y	&	Ib	&	NGC 1187	&	$	-			$	&	20.82	&	-	&	--	\\
2008D	&	Ib	&	NGC 2770	&	$	30.01	\pm	1.77	$	&	31.35	&	13	&	(1, 15-19)	\\
2009iz	&	Ib	&	UGC 02175	&	$	-			$	&	63.89	&	--	&	--	\\
2009jf	&	Ib	&	NGC 7479	&	$	33.58	\pm	3.34	$	&	35.59	&	4	&	(19)	\\
1994I	&	Ic	&	M51	&	$	7.63	\pm	0.07	$	&	6.91	&	30	&	(19-37)	\\
2002ap	&	Ic	&	M74	&	$	9.22	\pm	0.61	$	&	9.76	&	15	&	(19,26,44-51)	\\
2004aw	&	Ic	&	NGC 3997	&	$	78.11	\pm	11.76	$	&	78.99	&	2	&	(1)	\\
2007gr	&	Ic	&	NGC 1058	&	$	10.09	\pm	0.67	$	&	7.71	&	6	&	(16,19,38-42)	\\
2011bm	&	Ic	&	IC 3918	&	$	129.38	\pm	5.16	$	&	100.09	&	6	&	(3,7,9,43)	\\
2006T	&	IIb	&	NGC 3054	&	$	34.54	\pm	1.09	$	&	35.85	&	15	&	(1,13,15-17,19,52)	\\
2008ax	&	IIb	&	NGC 4490	&	$	7.88	\pm	0.72	$	&	9.37	&	7	&	(1,13,19,53)	\\
2008bo	&	IIb	&	NGC 6643	&	$	20.88	\pm	1.31	$	&	21.91	&	12	&	(1,15-19)	\\
2010as	&	IIb	&	NGC 6000	&	$	26.13	\pm	2.54	$	&	32.94	&	6	&	(1,13,19)	\\
2011dh	&	IIb	&	M51	&	$	7.63	\pm	0.07	$	&	6.91	&	30	&	(19-37)	\\
2011ei	&	IIb	&	NGC 6925	&	$	30.07	\pm	30.13	$	&	41.80	&	18	&	(1,15-19,52)	\\
2011fu	&	IIb	&	UGC 01626	&	$	-			$	&	83.52	&	--	&	--	\\
2011hs	&	IIb	&	IC 5267	&	$	27.61	\pm	3.29	$	&	25.51	&	5	&	(15,19)	\\
PTF11rks	&	SLSN-Ic	&	ANON	&	$	-			$	&	960.72	&	--	&	--	\\
2011ke	&	SLSN-Ic	&	ANON	&	$	-			$	&	702.94	&	--	&	--	\\
2012il	&	SLSN-Ic	&	ANON	&	$	-			$	&	877.11	&	--	&	--	\\
PTF12dam	&	SLSN-Ic	&	ANON	&	$	-			$	&	514.06	&	--	&	--	\\
2003jd	&	IcBL	&	MCG-01-59-21	&	$	-			$	&	85.06	&	--	&	--	\\
2007ru	&	IcBL	&	UGC 12381	&	$	-			$	&	69.68	&	--	&	--	\\
2009bb	&	IcBL	&	NGC 3278	&	$	40.68	\pm	4.36	$	&	44.83	&	4	&	(1)	\\
2012ap	&	IcBL	&	NGC 1729	&	$	40.37	\pm	7.38	$	&	55.04	&	2	&	(16)	\\
1998bw	&	GRB	&	ESO 184-G82	&	$	-			$	&	38.42	&	--	&	--	\\
2003dh	&	GRB	&	ANON	&	$	-			$	&	841.27	&	--	&	--	\\
2003lw	&	GRB	&	ANON	&	$	-			$	&	506.36	&	--	&	--	\\
2006aj	&	GRB	&	ANON	&	$	-			$	&	152.62	&	--	&	--	\\
2010bh	&	GRB	&	ANON	&	$	-			$	&	274.41	&	--	&	--	\\
2013dx	&	GRB	&	ANON	&	$	-			$	&	713.66	&	--	&	--	\\
\hline
\end{tabular}
\begin{flushleft}
$^{\dagger}$The weighted average linear distance and its associated error.\\
$^{\ddagger}$Luminosity distance calculated for a $\Lambda$CDM cosmology constrained by Planck (Planck Collaboration et al. 2013) of $H_{0} = 67.3$ km s$^{-1}$ Mpc$^{-1}$, $\Omega_{\rm M} = 0.315$, $\Omega_{\Lambda} = 0.685$.\\ 
\scriptsize{\textbf{References}: (1) Theureau et al. (2007); (2) Wood-Vasey et al. (2008); (3) Mandel et al. (2011); (4) Mandel et al. (2009); (5)~Wang et al. (2006); (6) Takanashi et al. (2008); (7) Ganeshalingam et al. (2013); (8) Weyant et al. (2014); (9) Hicken et al. (2009); (10) Prieto et al. (2006); (11) Jha et al. (2007); (12) Reindl et al. (2005); (13) Terry et  al. (2002); (14) Sorce et al. (2012); (15) Willick et al. (1997); (16) Springob et al. (2009); (17) Tully et al. (2009); (18) Tully et al. (1992); (19) Tully (1988); (20) Ciardullo et al. (2002); (21) Ferrarese et al. (2000); (22) Feldmeier et al. (1997); (23) Tonry et al. (2001); (24)~Richmond et al. (1996); (25) Sofue (1991); (26) Zasov  \& Bizyaev (1996); (27) Bose  \& Kumar (2014); (28) Tak{\'a}ts  \& Vink{\'o} (2006); (29) Iwamoto et al. (1994); (30) Tak{\'a}ts  \& Vink{\'o} (2012); (31) Baron et al. (2007); (32) Baron et al. (1996); (33)~Poznanski et al. (2009); (34) Vink{\'o} et  al. (2012); (35) Dessart et al. (2008); (36) Tutui  \& Sofue (1997); (37) Chiba  \& Yoshii (1995); (38) Schmidt et al. (1994); (39) Schmidt et al. (1992); (40) Kirshner  \& Kwan (1974); (41) Zinn et  al. (2011); (42) Pierce (1994); (43) Amanullah et al. (2010); (44) Sohn  \& Davidge (1996); (45) Sharina et  al. (1996); (46) Hendry et al. (2005); (47) Herrmann et al.(2008); (48) Vink{\'o} et  al. (2004); (49) Van Dyk et al. (2006); (50) Olivares E.~et al. (2010); (51)~Jang  \& Lee (2014); (52) Pedreros  \& Madore (1981); (53) Karachentsev et al. (2013).}\\
\end{flushleft}
\end{table*}

\begin{landscape}
\begin{table}
\centering
\caption{Luminosity--decline parameters of SNe Ib, Ic, IIb, SLSN-Ic, IcBL and GRB-SNe}
\setlength{\tabcolsep}{13pt}
\label{table:lum_decline}
\begin{tabular}{|cccccc|ccc|ccc|}
\hline
	&		&		&		&		&		&		MC				&		MC				&	MC	&		B				&		B				&	B	\\
\hline																																							
D$^{*}$	&	Type	&	Filter	&	$N$	&	$r$	&	$P$	&		$m$				&		$b$				&	$\sigma$	&		$m$				&		$b$				&	$\sigma$	\\
\hline																																							
$D_{L}$	&	Ib	&	$B$	&	11	&	0.043	&	0.901	&	$	0.08	\pm	0.11	$	&	$	-17.31	\pm	0.14	$	&	0.58	&	$	1.81	\pm	0.03	$	&	$	-19.14	\pm	0.03	$	&	0.80	\\
$D_{L}$	&	Ib	&	$V$	&	11	&	0.072	&	0.832	&	$	0.20	\pm	0.17	$	&	$	-17.72	\pm	0.13	$	&	0.55	&	$	2.76	\pm	0.05	$	&	$	-19.53	\pm	0.04	$	&	0.77	\\
$D_{L}$	&	Ib	&	$R$	&	11	&	0.099	&	0.772	&	$	0.38	\pm	0.28	$	&	$	-17.89	\pm	0.16	$	&	0.58	&	$	2.58	\pm	0.05	$	&	$	-18.96	\pm	0.03	$	&	0.69	\\
Lin.	&	Ib	&	$B$	&	8	&	0.148	&	0.726	&	$	0.43	\pm	0.18	$	&	$	-17.57	\pm	0.21	$	&	0.81	&	$	3.32	\pm	0.11	$	&	$	-21.09	\pm	0.14	$	&	1.17	\\
Lin.	&	Ib	&	$V$	&	8	&	0.207	&	0.623	&	$	0.99	\pm	0.37	$	&	$	-18.18	\pm	0.28	$	&	0.75	&	$	5.64	\pm	0.20	$	&	$	-22.23	\pm	0.18	$	&	1.21	\\
Lin.	&	Ib	&	$R$	&	8	&	0.253	&	0.546	&	$	1.51	\pm	0.73	$	&	$	-18.38	\pm	0.39	$	&	0.77	&	$	-5.03	\pm	0.17	$	&	$	-13.93	\pm	0.13	$	&	1.45	\\
$D_{L}$	&	Ic	&	$B$	&	5	&	0.398	&	0.507	&	$	0.44	\pm	0.14	$	&	$	-17.94	\pm	0.16	$	&	0.59	&	$	1.78	\pm	0.07	$	&	$	-19.23	\pm	0.07	$	&	1.08	\\
$D_{L}$	&	Ic	&	$V$	&	5	&	0.580	&	0.305	&	$	0.76	\pm	0.09	$	&	$	-18.46	\pm	0.06	$	&	0.52	&	$	2.15	\pm	0.06	$	&	$	-19.41	\pm	0.05	$	&	0.89	\\
$D_{L}$	&	Ic	&	$R$	&	5	&	0.584	&	0.301	&	$	0.91	\pm	0.09	$	&	$	-18.61	\pm	0.05	$	&	0.58	&	$	2.72	\pm	0.08	$	&	$	-19.38	\pm	0.05	$	&	1.14	\\
Lin.	&	Ic	&	$B$	&	5	&	0.478	&	0.416	&	$	0.55	\pm	0.16	$	&	$	-18.31	\pm	0.19	$	&	0.58	&	$	1.64	\pm	0.07	$	&	$	-19.38	\pm	0.07	$	&	0.94	\\
Lin.	&	Ic	&	$V$	&	5	&	0.633	&	0.215	&	$	0.85	\pm	0.10	$	&	$	-18.77	\pm	0.10	$	&	0.50	&	$	2.04	\pm	0.07	$	&	$	-19.75	\pm	0.06	$	&	0.77	\\
Lin.	&	Ic	&	$R$	&	5	&	0.652	&	0.235	&	$	1.05	\pm	0.11	$	&	$	-18.95	\pm	0.09	$	&	0.56	&	$	2.25	\pm	0.08	$	&	$	-19.75	\pm	0.06	$	&	0.79	\\
$D_{L}$	&	IIB	&	$B$	&	8	&	0.526	&	0.181	&	$	1.75	\pm	0.52	$	&	$	-19.51	\pm	0.79	$	&	0.76	&	$	5.19	\pm	0.23	$	&	$	-23.84	\pm	0.34	$	&	1.33	\\
$D_{L}$	&	IIB	&	$V$	&	7	&	0.694	&	0.084	&	$	2.21	\pm	0.33	$	&	$	-19.36	\pm	0.33	$	&	0.58	&	$	2.88	\pm	0.16	$	&	$	-19.41	\pm	0.14	$	&	0.84	\\
$D_{L}$	&	IIB	&	$R$	&	7	&	0.431	&	0.334	&	$	1.64	\pm	0.34	$	&	$	-18.54	\pm	0.23	$	&	0.72	&	$	2.43	\pm	0.06	$	&	$	-18.46	\pm	0.04	$	&	0.96	\\
Lin.	&	IIB	&	$B$	&	7	&	0.506	&	0.247	&	$	1.28	\pm	0.47	$	&	$	-18.51	\pm	0.72	$	&	0.62	&	$	4.09	\pm	0.24	$	&	$	-22.35	\pm	0.35	$	&	0.98	\\
Lin.	&	IIB	&	$V$	&	6	&	0.543	&	0.266	&	$	1.34	\pm	0.34	$	&	$	-18.19	\pm	0.34	$	&	0.54	&	$	2.54	\pm	0.22	$	&	$	-19.04	\pm	0.19	$	&	0.70	\\
Lin.	&	IIB	&	$R$	&	6	&	-0.047	&	0.930	&	$	-0.15	\pm	0.35	$	&	$	-16.94	\pm	0.25	$	&	0.68	&	$	-3.70	\pm	0.20	$	&	$	-14.55	\pm	0.13	$	&	1.01	\\
$D_{L}$	&	SLSN-Ic	&	$B$	&	4	&	0.406	&	0.594	&	$	0.86	\pm	0.43	$	&	$	-21.77	\pm	0.16	$	&	0.36	&	$	2.33	\pm	0.15	$	&	$	-22.11	\pm	0.03	$	&	0.58	\\
$D_{L}$	&	SLSN-Ic	&	$V$	&	3	&	-0.133	&	0.914	&	$	-0.17	\pm	0.60	$	&	$	-21.42	\pm	0.19	$	&	0.43	&	$	2.01	\pm	0.25	$	&	$	-21.86	\pm	0.04	$	&	0.71	\\
$D_{L}$	&	SLSN-Ic	&	$R$	&	3	&	0.999	&	0.024	&	$	1.89	\pm	0.22	$	&	$	-21.72	\pm	0.09	$	&	0.01	&	$	1.98	\pm	0.14	$	&	$	-21.75	\pm	0.03	$	&	0.01	\\
\hline
$D_{L}$	&	IcBL	&	$B$	&	4	&	0.600	&	0.400	&	$	1.10	\pm	0.95	$	&	$	-20.25	\pm	1.37	$	&	0.21	&	$	2.54	\pm	0.94	$	&	$	-22.44	\pm	1.35	$	&	0.34	\\
$D_{L}$	&	IcBL	&	$V$	&	4	&	0.494	&	0.506	&	$	0.98	\pm	0.76	$	&	$	-20.01	\pm	0.70	$	&	0.19	&	$	2.00	\pm	0.42	$	&	$	-20.90	\pm	0.40	$	&	0.23	\\
$D_{L}$	&	IcBL	&	$R$	&	4	&	0.773	&	0.227	&	$	2.79	\pm	1.11	$	&	$	-21.25	\pm	0.86	$	&	0.16	&	$	3.24	\pm	0.62	$	&	$	-21.55	\pm	0.50	$	&	0.17	\\
Lin.	&	IcBL	&	$B$	&	2	&	1.000	&	0.000	&	$	2.23	\pm	58.93	$	&	$	-21.12	\pm	74.94	$	&	0.14	&	$	2.69	\pm	2.11	$	&	$	-21.77	\pm	2.79	$	&	0.20	\\
Lin.	&	IcBL	&	$V$	&	2	&	1.000	&	0.000	&	$	0.79	\pm	1.87	$	&	$	-19.49	\pm	1.99	$	&	0.01	&	$	2.44	\pm	2.12	$	&	$	-21.29	\pm	2.29	$	&	0.21	\\
Lin.	&	IcBL	&	$R$	&	2	&	1.000	&	0.000	&	$	1.98	\pm	2.77	$	&	$	-20.27	\pm	2.37	$	&	0.02	&	$	4.36	\pm	2.61	$	&	$	-22.34	\pm	2.27	$	&	0.23	\\
$D_{L}$	&	GRB	&	$B$	&	4	&	0.988	&	0.012	&	$	1.17	\pm	0.67	$	&	$	-20.33	\pm	0.91	$	&	0.06	&	$	1.58	\pm	0.47	$	&	$	-20.87	\pm	0.63	$	&	0.09	\\
$D_{L}$	&	GRB	&	$V$	&	6	&	0.918	&	0.010	&	$	1.55	\pm	0.41	$	&	$	-20.63	\pm	0.37	$	&	0.20	&	$	1.78	\pm	0.30	$	&	$	-20.97	\pm	0.27	$	&	0.23	\\
$D_{L}$	&	GRB	&	$R$	&	5	&	0.730	&	0.162	&	$	2.12	\pm	0.76	$	&	$	-20.70	\pm	0.53	$	&	0.33	&	$	3.05	\pm	0.54	$	&	$	-21.36	\pm	0.37	$	&	0.34	\\
\hline
$D_{L}$	&	GRB+IcBL	&	$B$	&	8	&	0.861	&	0.006	&	$	1.06	\pm	0.47	$	&	$	-20.19	\pm	0.66	$	&	0.16	&	$	1.78	\pm	0.37	$	&	$	-21.32	\pm	0.51	$	&	0.24	\\
$D_{L}$	&	GRB+IcBL	&	$V$	&	10	&	0.871	&	0.001	&	$	1.48	\pm	0.36	$	&	$	-20.54	\pm	0.33	$	&	0.21	&	$	2.20	\pm	0.23	$	&	$	-21.14	\pm	0.22	$	&	0.25	\\
$D_{L}$	&	GRB+IcBL	&	$R$	&	9	&	0.743	&	0.022	&	$	2.10	\pm	0.59	$	&	$	-20.70	\pm	0.44	$	&	0.28	&	$	2.98	\pm	0.34	$	&	$	-21.34	\pm	0.26	$	&	0.28	\\
\hline																																																					
\end{tabular}
\begin{flushleft}
$^{*}$Luminosity distance ($D_{L}$) or linear distance (Lin.) -- see Table \ref{table:distances}.\\
MC: MC--LLS Bootstrap method; B: Bayesian--MultiNest method.\\
\end{flushleft}
\end{table}
\end{landscape}

\nopagebreak

\begin{table*}
\centering
\caption{Hubble-diagram offsets of SNe Ib, Ic, IIb, SLSN-Ic, IcBL and GRB-SNe (uncorrected for a luminosity--decline relationship)}
\setlength{\tabcolsep}{15pt}
\label{table:hubble_offsets_uncorrected}
\begin{tabular}{|ccc|cc|cc|cc|}
\hline
	&		&		&		MC				&	MC	&		B				&	B	\\
\hline																					
Type	&	Filter	&	$N$	&		$b$				&	$\sigma$	&		$b$				&	$\sigma$	\\
\hline																					
Ia ($z\le0.2$)	&	$B$	&	318	&	$	0.631	\pm	0.001	$	&	0.30	&	$	0.631	\pm	5.58E-05	$	&	0.30	\\
Ia ($z\le0.1$)	&	$B$	&	152	&	$	0.643	\pm	0.001	$	&	0.31	&	$	0.643	\pm	8.09E-05	$	&	0.31	\\
Ib	&	$B$	&	11	&	$	0.286	\pm	0.008	$	&	0.63	&	$	0.335	\pm	9.37E-04	$	&	0.68	\\
Ib	&	$V$	&	11	&	$	0.357	\pm	0.007	$	&	0.59	&	$	0.398	\pm	8.86E-04	$	&	0.63	\\
Ib	&	$R$	&	11	&	$	0.381	\pm	0.006	$	&	0.63	&	$	0.419	\pm	8.72E-04	$	&	0.65	\\
Ic	&	$B$	&	5	&	$	0.304	\pm	0.008	$	&	0.64	&	$	0.418	\pm	1.50E-03	$	&	0.85	\\
Ic	&	$V$	&	5	&	$	0.386	\pm	0.007	$	&	0.63	&	$	0.502	\pm	1.47E-03	$	&	0.86	\\
Ic	&	$R$	&	5	&	$	0.415	\pm	0.007	$	&	0.70	&	$	0.542	\pm	1.44E-03	$	&	0.95	\\
IIB	&	$B$	&	8	&	$	0.211	\pm	0.012	$	&	0.88	&	$	0.371	\pm	8.25E-04	$	&	1.19	\\
IIB	&	$V$	&	7	&	$	0.276	\pm	0.010	$	&	0.79	&	$	0.454	\pm	8.50E-04	$	&	1.19	\\
IIB	&	$R$	&	7	&	$	0.314	\pm	0.009	$	&	0.79	&	$	0.485	\pm	8.40E-04	$	&	1.17	\\
SLSN-Ic	&	$B$	&	4	&	$	1.060	\pm	0.010	$	&	0.43	&	$	1.030	\pm	1.22E-04	$	&	0.46	\\
SLSN-Ic	&	$V$	&	3	&	$	1.082	\pm	0.011	$	&	0.47	&	$	1.030	\pm	1.52E-04	$	&	0.52	\\
SLSN-Ic	&	$R$	&	3	&	$	1.040	\pm	0.008	$	&	0.33	&	$	1.000	\pm	1.56E-04	$	&	0.37	\\
IcBL	&	$B$	&	2	&	$	0.570	\pm	0.017	$	&	0.27	&	$	0.577	\pm	1.49E-03	$	&	0.27	\\
IcBL	&	$V$	&	2	&	$	0.638	\pm	0.014	$	&	0.22	&	$	0.636	\pm	1.49E-03	$	&	0.22	\\
IcBL	&	$R$	&	2	&	$	0.639	\pm	0.012	$	&	0.26	&	$	0.641	\pm	1.49E-03	$	&	0.26	\\
GRB	&	$B$	&	4	&	$	0.559	\pm	0.022	$	&	0.31	&	$	0.525	\pm	2.58E-04	$	&	0.35	\\
GRB	&	$V$	&	6	&	$	0.652	\pm	0.015	$	&	0.46	&	$	0.665	\pm	1.88E-04	$	&	0.47	\\
GRB	&	$R$	&	5	&	$	0.654	\pm	0.013	$	&	0.46	&	$	0.671	\pm	2.00E-04	$	&	0.47	\\
\hline
GRB+IcBL	&	$B$	&	8	&	$	0.564	\pm	0.014	$	&	0.29	&	$	0.527	\pm	2.52E-04	$	&	0.34	\\
GRB+IcBL	&	$V$	&	10	&	$	0.647	\pm	0.011	$	&	0.39	&	$	0.665	\pm	1.88E-04	$	&	0.40	\\
GRB+IcBL	&	$R$	&	9	&	$	0.647	\pm	0.009	$	&	0.38	&	$	0.670	\pm	1.99E-04	$	&	0.40	\\
\hline																													
\end{tabular}
\begin{flushleft}
MC: MC--LLS Bootstrap method; B: Bayesian--MultiNest method.\\
\end{flushleft}
\end{table*}

\begin{table*}
\centering
\caption{Value of the Hubble constant derived from SNe Ib, Ic, IIb, SLSN-Ic, IcBL and GRB-SNe}
\setlength{\tabcolsep}{11pt}
\label{table:H0_per_SN_per_filter}
\begin{tabular}{|cccc|ccc|ccc|}
\hline
	&		&		&		&		MC						&		MC		&	MC	&		B				&	B	&	B	\\
		\hline																													
	&	Type	&	Filter	&	$N$	&		$H_{0,\rm w}$						&		$\bar{H}_{0}$		&	st.dev.	&		$H_{0,\rm w}$				&	$\bar{H}_{0}$	&	st.dev.	\\
		\hline																													
Uncorr.	&	Ib	&	$B$	&	8	&	$	72.2	\pm	2.1		$		&		79.6		&	31.9	&	$	83.0	\pm	1.3	$	&	89.1	&	35.7	\\
Uncorr.	&	Ib	&	$V$	&	8	&	$	70.4	\pm	1.9		$		&		78.1		&	29.9	&	$	78.5	\pm	1.2	$	&	85.9	&	32.8	\\
Uncorr.	&	Ib	&	$R$	&	8	&	$	69.8	\pm	1.5		$		&		78.9		&	32.2	&	$	76.7	\pm	0.9	$	&	86.1	&	35.1	\\
Uncorr.	&	Ic	&	$B$	&	5	&	$	46.9	\pm	2.2		$		&		62.5		&	19.4	&	$	59.9	\pm	2.3	$	&	81.3	&	25.2	\\
Uncorr.	&	Ic	&	$V$	&	5	&	$	44.2	\pm	1.9		$		&		62.4		&	18.1	&	$	56.3	\pm	2.0	$	&	81.5	&	23.6	\\
Uncorr.	&	Ic	&	$R$	&	5	&	$	41.7	\pm	1.8		$		&		63.1		&	20.1	&	$	54.4	\pm	1.9	$	&	84.6	&	26.9	\\
Uncorr.	&	IIB	&	$B$	&	7	&	$	87.8	\pm	3.1		$		&		82.3		&	26.6	&	$	134.8	\pm	1.9	$	&	118.9	&	38.4	\\
Uncorr.	&	IIB	&	$V$	&	6	&	$	80.7	\pm	3.1		$		&		83.3		&	25.5	&	$	122.0	\pm	2.9	$	&	125.5	&	38.4	\\
Uncorr.	&	IIB	&	$R$	&	6	&	$	77.5	\pm	2.3		$		&		84.2		&	28.6	&	$	112.7	\pm	1.6	$	&	124.8	&	42.4	\\
Uncorr.	&	IcBL	&	$B$	&	2	&	$	84.3	\pm	15.5		$		&		82.6		&	7.5	&	$	86.0	\pm	13.0	$	&	83.9	&	7.6	\\
Uncorr.	&	IcBL	&	$V$	&	2	&	$	80.3	\pm	14.0		$		&		79.7		&	3.9	&	$	80.0	\pm	11.9	$	&	79.3	&	3.9	\\
Uncorr.	&	IcBL	&	$R$	&	2	&	$	83.0	\pm	12.8		$		&		81.8		&	6.7	&	$	83.6	\pm	11.1	$	&	82.2	&	6.7	\\
\hline
Uncorr.	&	GRB+IcBL	&	$B$	&	2	&	$	83.2	\pm	14.8		$		&		81.5		&	7.4	&	$	76.7	\pm	11.3	$	&	74.8	&	6.8	\\
Uncorr.	&	GRB+IcBL	&	$V$	&	2	&	$	82.0	\pm	13.8		$		&		81.3		&	4.0	&	$	85.6	\pm	12.5	$	&	84.8	&	4.1	\\
Uncorr.	&	GRB+IcBL	&	$R$	&	2	&	$	84.6	\pm	12.5		$		&		83.3		&	6.8	&	$	89.4	\pm	11.6	$	&	87.9	&	7.1	\\
\hline																																																					
\end{tabular}
\begin{flushleft}
Uncorrected (Uncorr.) for a luminosity--decline relationship (linear distance only).\\
The weighted ($H_{0,\rm w}$) and average ($\bar{H}_{0}$) values of the Hubble constant are in units of km~s$^{-1}$~Mpc$^{-1}$.\\
MC: MC--LLS Bootstrap method; B: Bayesian--MultiNest method.\\
\end{flushleft}
\end{table*}

\begin{table*}
\centering
\caption{Summary of $H_{0}$ derived from SNe Ib, Ic, IIb, SLSN-Ic, IcBL and GRB-SNe}
\setlength{\tabcolsep}{17pt}
\label{table:H0_summary}
\begin{tabular}{|cc|c|c|c|c|}
\hline
& Method	&	Type	&		$H_{0,\rm w}$				&	$\bar{H}_{0}$	&	st.dev.	\\
\hline
Uncorr.	&	MC	&	Ib	&	$	70.6	\pm	1.0	$	&		78.9		&		31.3		\\
Uncorr.	&	B	&	Ib	&	$	78.7	\pm	0.6	$	&		87.0		&		34.6		\\
Uncorr.	&	MC	&	Ic	&	$	43.9	\pm	1.1	$	&		62.7		&		19.2		\\
Uncorr.	&	B	&	Ic	&	$	56.5	\pm	1.2	$	&		82.5		&		25.2		\\
Uncorr.	&	MC	&	IIB	&	$	81.0	\pm	1.6	$	&		83.3		&		26.9		\\
Uncorr.	&	B	&	IIB	&	$	121.8	\pm	1.1	$	&		123.1		&		39.8		\\
\hline
Uncorr.	&	MC	&	IcBL	&	$	82.5	\pm	8.1	$	&		81.4		&		6.0		\\
Uncorr.	&	B	&	IcBL	&	$	83.1	\pm	6.9	$	&		81.8		&		6.1		\\
\hline																													
\end{tabular}
\begin{flushleft}
Uncorrected (Uncorr.) for a luminosity--decline relationship (linear distance only).\\
The weighted ($H_{0,\rm w}$) and average ($\bar{H}_{0}$) values of the Hubble constant are in units of km~s$^{-1}$~Mpc$^{-1}$.\\
MC: MC--LLS Bootstrap method; B: Bayesian--MultiNest method.\\
\end{flushleft}
\end{table*}


\label{lastpage}

\end{document}